\documentclass[sor,jor,reprint,amsmath,amssymb,superscriptaddress,longbibliography]{revtex4-2}
\usepackage{graphicx}
\usepackage{bm}
\usepackage{booktabs}
\usepackage{xcolor}
\usepackage[hidelinks]{hyperref}
\newcommand{\rank}{\operatorname{rank}}
\newcommand{\diag}{\operatorname{diag}}
\newcommand{\dev}{\operatorname{dev}}
\newcommand{\tr}{\operatorname{tr}}

\newcommand{\calD}{\mathcal{D}}
\hypersetup{pdftitle={Calibrated shape-only recovery of membrane viscosities},pdfauthor={Julian Olszewski, Zijian Liu, Jie Xu}}
\begin{document}
\title{Calibrated shape-only recovery of membrane viscosities}
\author{Julian Olszewski}
\thanks{These authors contributed equally to this work.}
\affiliation{Department of Mechanical and Industrial Engineering,
\mbox{University of Illinois Chicago, Chicago, Illinois 60607, USA}}
\author{Zijian Liu}
\thanks{These authors contributed equally to this work.}
\affiliation{Department of Mechanical and Industrial Engineering,
\mbox{University of Illinois Chicago, Chicago, Illinois 60607, USA}}
\author{Jie Xu}
\email{jiexu@uic.edu}
\affiliation{Department of Mechanical and Industrial Engineering,
\mbox{University of Illinois Chicago, Chicago, Illinois 60607, USA}}
\date{October 16, 2026}
\begin{abstract}
Nearby relaxation rates make two membrane viscosities difficult to
distinguish from a shape transient. We show how calibrated preparation
removes this loss of local sensitivity in a linear Kelvin--Voigt sphere
with finite area elasticity and intrinsic membrane dissipation.
Two independent releases, with a known signed amplitude ratio and
calibrated relative gain and timing, yield normal-displacement records
whose fixed linear combination separates the shear and dilatational
rates. Their common amplitude remains unknown. For fixed sampling,
positive-definite weights, bounded positive rates and nonzero excitation,
the amplitude-projected sensitivity has a positive lower bound even at
rate coincidence. Without relative calibration, the paired sensitivity
decreases linearly with the rate gap; the single tangential-release
record decreases quadratically. The calibrated bound is compatible
with fixed finite initial energy. A positive local bound also persists
with independent ratio calibration of fixed positive precision under
a Gaussian error model. Supplied elastic coefficients
convert the recovered rates into the two surface viscosities. An
unbounded spherical Stokes extension quantifies how bulk dissipation
changes the modal patterns and provides a sufficient full-Jacobian
criterion for retaining local material sensitivity.
\end{abstract}
\maketitle
\enlargethispage{-\baselineskip}

\section{Introduction}
Surface shear and dilatational viscosities describe distinct dissipative
responses of a membrane, but they are difficult to measure separately.
Representative capsule and red-blood-cell simulations set the two
coefficients equal and use one membrane-viscosity parameter \citep{Guglietta2021,Guglietta2023}. Release-and-relax
measurements likewise commonly extract one timescale: examples include
a vesicle after a field is removed \citep{Zhang2013} and a
capsule leaving a microfluidic expansion \citep{Gires2016}. A second
coefficient requires an excitation and an observation that distinguish
the two dissipative responses.

Small-deformation capsule theory and simulations describe membrane
viscosity in several flow regimes \citep{Barthes1985,Yazdani2013,Lu2023}.
Viscoelastic shell theory resolves spherical shear, compression and
bending modes, including solvent coupling
\citep{Rochal2005,Kuriabova2008,KuriabovaErratum}. Related work treats
interfacial viscosity in droplets \citep{Guglietta2026} and independent
surface viscosities in fluid-bilayer fluctuations \citep{Vlahovska2026}.
\citet{RegazziLeonetti2026} analyze shear-driven capsules with independent
surface viscosities and identify deformation and orientation as
complementary observables for their recovery under specified elastic
laws. These studies establish both the forward role of the two
coefficients and routes to shape-based inference. For a release-and-relax
measurement, the remaining design question is how prescribed initial
states distinguish two decay rates as those rates approach one another.

\citet{Panda2026} distinguish surface-viscous
effects through nonlinear Faraday-wave patterns. For their initially
flat Newtonian interface, the linear threshold and initial growth
depend on the sum of the viscosities, whereas nonlinear oblique modes
respond differently to shear and dilatational dissipation. The
measurement therefore has to retain distinct responses to the two
coefficients in its particular mechanical regime.

Inverse identification of elastic capsule properties is established
\citep{Wang2021}. For viscous relaxation, recovering two coefficients
requires resolving two rates even when their time scales are similar.
Structural identifiability asks which parameter changes can leave a
record unchanged \citep{Jacquez1985,Eisenberg2014}. Prescribed initial
conditions can create or remove such ambiguities
\citep{Saccomani2003,Joubert2021}, and identification from multiple
initial conditions is established for linear systems \citep{Guo2010}.
The issue addressed here is how preparation calibration changes the
local sensitivity of that recovery as the two rates coincide.

We use an intrinsic, zero-bending Kelvin--Voigt sphere with
Boussinesq--Scriven surface viscosity \citep{Boussinesq1913,Scriven1960},
the corresponding limit of the shell description of \citet{Rochal2005}.
Its shear elasticity and finite area elasticity permit both dissipation
mechanisms. This is distinct from the area-incompressible fluid-bilayer
model used for vesicle shear microrheometry \citep{Faizi2022}.
The intrinsic relaxation rates are $r_T=G_s/\mu_s$ and
$r_D=G_sc/\mu_s'$, where $c=1+2C_a$ measures area elasticity.
The commonly imposed equality $\mu_s'=\mu_s$ gives $r_D=cr_T$;
rate coincidence instead requires $\mu_s'=c\mu_s$.
Rate coincidence thus tests the observation design at a specific
material relation, rather than at equal surface viscosities.

Our central result is that relative calibration of two independent
releases preserves positive local two-rate sensitivity through
coincidence while their common amplitude is fitted. A radial release
provides a dilatational reference that can be combined with the normal
record of a tangential release to isolate the shear response.
Here shape-only describes the relaxation measurement; preparation
independently supplies the release fields, signed amplitude ratio,
relative observation gain and time origin. We derive the required
holding tractions and establish the sensitivity bound at finite
initial energy. Comparing calibrated, uncalibrated and single-record
measurements then identifies what calibration contributes. Finally,
an explicit unbounded Stokes extension gives a full-material-Jacobian
criterion for carrying this sensitivity into a fluid-coupled model.

\begin{figure*}[t]
 \includegraphics[width=\textwidth]{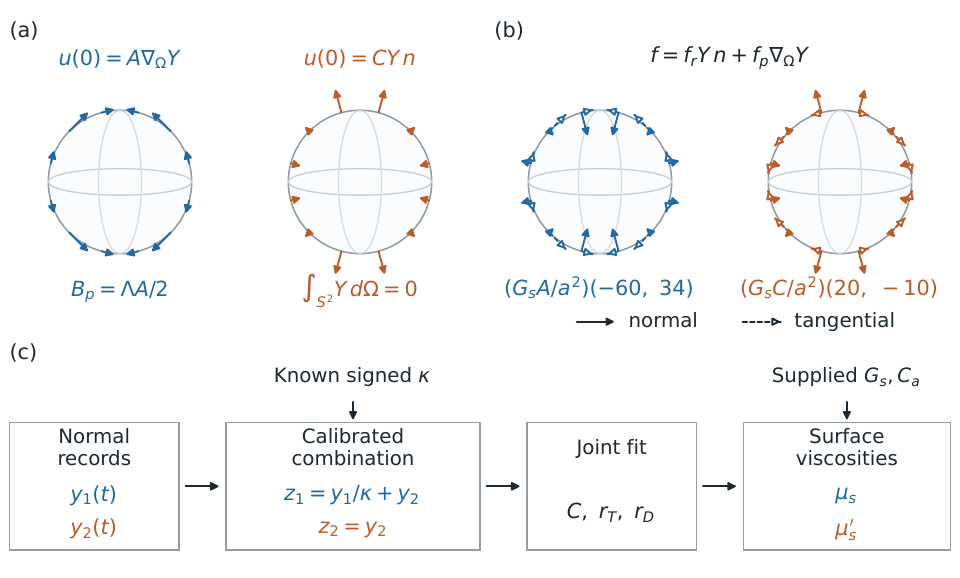}
 \caption{Prescribed releases and calibrated shape-only recovery in the
 reference membrane model. Blue marks the tangential-poloidal release and
 its record, orange the radial release and its record. (a) Meridional
 displacement arrows for the two releases, with normalized $Y=Y_{20}$,
 $\Lambda=6$, positive $A,C$ and $\nabla_\Omega$ the gradient on the unit
 sphere. The radial field expands and contracts locally with zero mean.
 (b) Holding tractions at $C_a=2$: arrows with filled heads show the
 normal component and dashed arrows with open heads the tangential
 component. The coefficient pairs are $(f_r,f_p)$. Arrow lengths are
 rescaled between drawings, with a common component scale within each
 traction drawing. (c) The normal records $y_1$ and $y_2$ of the two
 releases have signed amplitudes $B_p=\Lambda A/2$ and $C$. The known
 signed ratio $\kappa=B_p/C$ specifies the linear combination of the
 normal records shown in panel (c). The drawing specifies the
prescribed loading patterns and the reference-model signal map.}
 \label{fig:protocol}
\end{figure*}

Figure~\ref{fig:protocol} summarizes the prescribed preparation in the reference model and the
normal-record combination that underlies the analysis.

\section{Two resolved modes of a linear spherical membrane}
\subsection{Constitutive model and constraints}\label{sec:model}
Consider a sphere $S_a$ of radius $a$, unit normal $n$, and tangent
projector $\mathcal P=I-nn^{\mathsf T}$. A small material displacement is
$u=u_t+u_n n$. Its linear surface strain is
\begin{equation}
 E(u)=\operatorname{sym}\nabla_s u_t+\frac{u_n}{a}\mathcal P,
 \qquad
 \dev E=E-\frac{\tr E}{2}\mathcal P.
 \label{eq:strain}
\end{equation}
Here $\nabla_s u_t=\mathcal P(\nabla u_t)\mathcal P$ denotes the tangential
covariant gradient, so the strain is a tangential tensor.
For a shear modulus $G_s>0$ and finite area parameter $C_a\ge0$,
take the quadratic energy density
\begin{equation}
 w_2=G_s\left[\dev E:\dev E+
       \left(\frac12+C_a\right)(\tr E)^2\right].
 \label{eq:energy}
\end{equation}
This is the linear form of the shear-plus-area elastic law
\citep{Skalak1973,Yazdani2013}. Write the bilinear forms
\begin{align}
 R_s(u,v)&=2\int_{S_a}\dev E(u):\dev E(v)\,dA,\label{eq:formshear}\\
 R_d(u,v)&=\int_{S_a}\tr E(u)\tr E(v)\,dA .
 \label{eq:forms}
\end{align}
The surface-viscous stress uses positive coefficients
$\mu_s,\mu_s'$ in the deviatoric and dilatational parts,
$\tau^v=2\mu_s\,\dev D+\mu_s'(\tr D)\mathcal P$ with $D=E(\dot u)$, consistent
with the constitutive decomposition of \citet{Yazdani2013}
and \citet{Lu2023}. The resulting stiffness and
viscous resistance are
\begin{align}
 K&=G_s\bigl[R_s+(1+2C_a)R_d\bigr],\label{eq:stiffness}\\
 V&=\mu_s R_s+\mu_s'R_d .
 \label{eq:kv}
\end{align}
Dimensionally, $G_s/\mu_s$ and $G_s/\mu_s'$ are rates, with units of
inverse seconds. Table~\ref{tab:notation} collects the symbols used
throughout and points at the section that defines each.

\begin{table}[b]
\caption{Core symbols, grouped by their role. Specialized fluid matrices
are defined in Sec.~\ref{sec:fluid}; variants of $M$ specify which
amplitudes are fitted.}
\label{tab:notation}
\begin{ruledtabular}
\begin{tabular}{llc}
Symbol & Meaning & Defined in\\
\hline
\multicolumn{3}{l}{Geometry and material properties}\\
$a$ & sphere radius & \ref{sec:model}\\
$\ell$, $\Lambda=\ell(\ell+1)$ & harmonic degree and eigenvalue & \ref{sec:eigenfamilies}\\
$\phi_T$, $\phi_D$ & toroidal and radial eigenfields & \ref{sec:eigenfamilies}\\
$\psi_S$ & normal-visible, trace-free eigenfield & \ref{sec:releases}\\
$G_s$, $C_a$, $c=1+2C_a$ & shear modulus and area parameter & \ref{sec:model}\\
$\mu_s$, $\mu_s'$ & shear and dilatational surface viscosity & \ref{sec:model}\\
$\eta_{\rm in}$, $\eta_{\rm out}$ & inner and outer fluid viscosity & \ref{sec:fluid}\\
\hline
\multicolumn{3}{l}{Preparation, observation and calibration}\\
$A$, $C$ & tangential and radial release amplitudes & \ref{sec:releases}\\
$B_p=\Lambda A/2$ & signed amplitude in the first normal record & \ref{sec:separation}\\
$\kappa=B_p/C$ & known signed relative amplitude & \ref{sec:separation}\\
$(P,Q)$, $\mathcal M$ & release pattern and paired pattern matrix & \ref{sec:independence}\\
$S$ & common unknown record scale & \ref{sec:independence}\\
$r_T$, $r_D$ & shear and dilatational relaxation rates & \ref{sec:eigenfamilies}\\
\hline
\multicolumn{3}{l}{Mathematical analysis}\\
$R_s$, $R_d$ & shear and dilatational strain forms & \ref{sec:model}\\
$K$, $V$ & elastic stiffness and viscous resistance & \ref{sec:model}\\
$m$, $d$ & mean rate and half-gap & \ref{sec:separation}\\
$\Delta=r_T-r_D$ & full signed rate gap & \ref{sec:uncalibrated}\\
$\Delta_\Omega$ & unit-sphere Laplacian & \ref{sec:releases}\\
$W=L^{\mathsf T}L$ & fixed positive-definite observation weight & \ref{sec:sensitivity}\\
$J$, $N$ & target and fitted-nuisance derivatives & \ref{sec:sensitivity}\\
$M=P_{LN}^{\perp}LJ$ & sensitivity after fitted-amplitude adjustment & \ref{sec:sensitivity}\\
$\sigma_{\min}$, $\operatorname{cond}$ & smallest singular value and condition number & \ref{sec:sensitivity}
\end{tabular}
\end{ruledtabular}
\end{table}

The overdamped reference-membrane problem is
\begin{equation}
 V(\dot u,v)+K(u,v)=0
 \label{eq:weak}
\end{equation}
for admissible test displacements. We remove infinitesimal rigid
motions and enforce the linear volume condition
$\int_{S_a}u_n\,dA=0$. The area elasticity is finite, so
local dilatation remains admissible. Equation~(\ref{eq:weak})
defines the intrinsic, zero-bending membrane relaxation problem in
linear reference geometry; its rates describe that constitutive setting.
On the $\ell\ge2$ poloidal and toroidal spaces, $R_s$ and $R_d$
have no common nonzero kernel. The stated positive coefficients
therefore make $K$ and $V$ positive definite on these spaces. Testing with $\dot u$ gives
$\tfrac12\,dK(u,u)/dt=-V(\dot u,\dot u)\le0$, consistent with
the positive decay rates below.

\subsection{Analytically prescribed eigenfamilies}\label{sec:eigenfamilies}
Let $Y_{\ell m}$ be a real spherical harmonic normalized by
$\int_{S^2}Y_{\ell m}^2\,d\Omega=1$, let
$\Lambda=\ell(\ell+1)$, and write $\nabla_\Omega$ for the gradient on
the unit sphere. For $\ell\ge2$ define the dimensionless
displacement shapes
\begin{equation}
 \phi_T=n\times\nabla_\Omega Y_{\ell m},
 \qquad
 \phi_D=Y_{\ell m}n .
 \label{eq:modes}
\end{equation}
A scalar amplitude with dimensions of length multiplies each
shape. The toroidal field is divergence free and therefore has
$\tr E(\phi_T)=0$. The radial field has
$E(\phi_D)=(Y_{\ell m}/a)\mathcal P$ and therefore
$\dev E(\phi_D)=0$. Both are orthogonal to the rigid motions and
satisfy the linear volume condition. In particular the radial mode
has zero mean and is distinct from uniform dilation.

The normalized quadratic forms, derived in the supplementary
material, are
\begin{align}
 R_s(\phi_T,\phi_T)&=\Lambda(\Lambda-2),&
 R_d(\phi_T,\phi_T)&=0,\label{eq:toroidalnorm}\\
 R_s(\phi_D,\phi_D)&=0,&
 R_d(\phi_D,\phi_D)&=4.
 \label{eq:norms}
\end{align}
For every admissible test field, the zero trace or zero deviatoric
strain annihilates the corresponding bilinear form. Hence
$u_i(t)=A_i e^{-r_i t}\phi_i$ solves
Eq.~(\ref{eq:weak}) exactly with
\begin{equation}
 r_T=\frac{G_s}{\mu_s},\qquad
 r_D=\frac{G_s(1+2C_a)}{\mu_s'}.
 \label{eq:rates}
\end{equation}
The weak form thus fixes the two relaxation rates directly from the
shear and area strain forms.

\subsection{The two shape-only releases}\label{sec:releases}
The shape-only protocol uses a tangential displacement different from
the toroidal field $\phi_T$. The latter moves material along the
surface with zero normal displacement. Instead, define the
normal-visible, trace-free poloidal field
\begin{equation}
 \psi_S=\nabla_\Omega Y_{\ell m}
             +\frac{\Lambda}{2}Y_{\ell m}n,\qquad \ell\ge2.
\end{equation}
Since $\Delta_\Omega Y_{\ell m}=-\Lambda Y_{\ell m}$,
Eq.~(\ref{eq:strain}) gives $\tr E(\psi_S)=0$.
Its nonzero shear form is $\Lambda(\Lambda-2)$, so it relaxes
at $r_T$. For a purely tangential initial displacement
$u(0)=A\nabla_\Omega Y_{\ell m}
      =A\psi_S-(\Lambda A/2)\phi_D$, the normal
harmonic projection is
\begin{equation}
 a_n(t)=\frac{\Lambda A}{2}
          \left(e^{-r_Tt}-e^{-r_Dt}\right).
 \label{eq:poloidal}
\end{equation}
Both component fields satisfy the linear volume constraint.
For distinct rates and nonzero $A$, four equally spaced noiseless
samples determine
the unordered rate pair by standard exponential-sum reconstruction
\citep{Potts2010}, detailed in the supplementary material.
For this equal-and-opposite response, three distinct positive times
already suffice by Sec.~\ref{sec:collision}.
A separate radial release $u(0)=C\phi_D$ produces only the
dilatational exponential. It will supply the reference needed
to separate the two responses, as shown in Fig.~\ref{fig:protocol}.

Known signed initial amplitude labels the positive and negative
terms in Eq.~(\ref{eq:poloidal}); with unknown signed amplitude
the rate exchange is accompanied by $A\mapsto-A$.
At $r_T=r_D$ the normal signal is identically zero for every
common rate. The information carried by shape therefore depends
on the prescribed release.

\subsection{The traction that holds a prescribed release}
\label{sec:traction}
A prescribed initial displacement is prepared by an elastic holding
traction. Exciting and predicting a selected interface mode is an
established problem, for a driven meniscus \citep{XuAttinger2007} and for a
trapped gas-liquid interface whose modal spectrum is computed in advance
\citep{Gritsenko2018}. At equilibrium the velocity vanishes, so the
holding problem involves only the elastic stiffness. A finite hold
approximates this state once its slowest transient has decayed to the
chosen accuracy. The equilibrium satisfies
\begin{equation}
 K(u_{\rm hold},v)=\int f\cdot v\,dA
 \quad\text{for all admissible }v .
 \label{eq:holdproblem}
\end{equation}
The same $K$ determines the held states in both the intrinsic membrane
and fluid-coupled models.

Write a poloidal displacement as $u=q_rYn+q_p\nabla_\Omega Y$,
with length coordinates $q=(q_r,q_p)^{\mathsf T}$.
Pairing Eq.~(\ref{eq:holdproblem}) against a poloidal test field and
using $\int_{S^2}Y^2\,d\Omega=1$ and
$\int_{S^2}|\nabla_\Omega Y|^2\,d\Omega=\Lambda$ gives the normal
and tangential traction coefficients
\begin{equation}
 f_r=\frac{(Kq)_1}{a^2},
 \qquad
 f_p=\frac{(Kq)_2}{a^2\Lambda}.
 \label{eq:tractioncoefficients}
\end{equation}
At $\ell=2$ and $C_a=2$, for which
$K=G_s\bigl(\begin{smallmatrix}20&-60\\-60&204\end{smallmatrix}\bigr)$,
the two prescribed releases require
\begin{align}
 u&=C\,\phi_D:
 &(f_r,f_p)&=\frac{G_sC}{a^2}(20,-10),\nonumber\\
 u&=A\,\nabla_\Omega Y:
 &(f_r,f_p)&=\frac{G_sA}{a^2}(-60,34).
 \label{eq:tractionvalues}
\end{align}

Both releases require mixed normal and tangential holding tractions.
The radial release dilates area, and the resulting tension varies as
$Y_{\ell m}$, so its surface gradient must be balanced by an external
shear traction. For the tangential release,
$\nabla_\Omega Y$ has nonzero surface divergence,
$\mathrm{div}_\Omega\nabla_\Omega Y=-\Lambda Y$, so it too changes
local area, and on a curved surface that tension pushes normally. The
ratio of area-rms normal to area-rms tangential traction,
$|f_r|/(\sqrt{\Lambda}|f_p|)$, is $0.82$ for the radial release and
$0.72$ for the tangential one. Section III considers alternative
release patterns prepared by purely normal or tangential traction.

For $\ell\ge2$ the required loading has no rigid-body content, so it
carries zero net force and zero net torque and implies no external
anchoring. The area root-mean-square traction satisfies
$f_{\rm rms}^2=(4\pi)^{-1}\int_{S^2}|f|^2\,d\Omega
=(f_r^2+\Lambda f_p^2)/(4\pi)$.
At the reference scale $G_s=5\times10^{-6}$~N\,m$^{-1}$ and
$a=3.5$~$\mu$m, with $Y=Y_{20}$ and a peak areal strain of one percent,
the radial release needs an amplitude of $27.7$~nm and
$f_{\rm rms}=0.101$~Pa, and the tangential release $9.2$~nm and
$0.109$~Pa. The corresponding scale forces $4\pi a^2f_{\rm rms}$ are
$15.6$ and $16.8$~pN. These are elastic loading scales for the prescribed
states; the relative importance of membrane and fluid dissipation during
relaxation is determined separately in Sec.~\ref{sec:dominance}.

The loading also determines the signed preparation ratio
$\kappa=\Lambda A/(2C)$ used in Sec.~\ref{sec:separation}.
At the parameters of Eq.~(\ref{eq:tractionvalues}), write the applied
traction coefficients as $\alpha_T(-60,34)$ and $\alpha_D(20,-10)$.
These fixed patterns contain no $G_s$. Their coefficients satisfy
$\alpha_T=G_sA/a^2$ and $\alpha_D=G_sC/a^2$, so the common
compliance factor cancels and
\begin{equation}
 \kappa=\frac{\Lambda\alpha_T}{2\alpha_D}.
 \label{eq:kappafromactuator}
\end{equation}
Here $\alpha_T/\alpha_D$ is the signed coefficient ratio in the
specified patterns. Using force magnitudes instead requires the
corresponding pattern norms and signs. At any chosen harmonic the
patterns follow from $K/G_s$ and require the supplied area parameter
$c=1+2C_a$, independently of $G_s$ and the surface viscosities.
Thus relative loading determines $\kappa$ without determining $G_s$.
The preparation calibration supplies the realized release fields and
their signed amplitudes independently of the subsequent normal records.

The supplied area parameter also controls the release composition.
If tractions are designed with $c$ but the true parameter is $c'>0$,
application of Eq.~(\ref{eq:poloidalmatrices}) gives
\begin{align}
 q_{\rm rad}&=(cC/c',0)^{\mathsf T},\nonumber\\
 q_{\rm tan}&=\bigl[\tfrac{\Lambda A}{2}(1-c/c'),A\bigr]^{\mathsf T}.
 \label{eq:areaerror}
\end{align}
The tangential state now has an initial normal component. Its modal
coefficients are $(B_p,-B_pc/c')$; the radial coefficient is $Cc/c'$.
The nominal $\kappa=B_p/C$ still gives
$y_1/\kappa+y_2=C e^{-r_Tt}$ and $y_2=(c/c')C e^{-r_Dt}$.
The two rates remain separable by endpoint log-ratios, with unequal
transformed amplitudes. A statistical fit must therefore model $c/c'$
or fit these amplitudes separately. Absolute viscosity recovery also
uses the elastic calibration: with $G_s$ correct, substituting the
nominal $c$ gives $\widehat{\mu_s'}=(c/c')\mu_s'$.
An uncertain area parameter consequently belongs in the preparation
and material map, rather than in the scalar-ratio calibration alone.

\subsection{Resolved observations and elastic calibration}\label{sec:calibration}
For comparison with normal-only imaging, consider resolved material
displacements in the toroidal/radial eigenfamilies. Prescribe
$u(0)=A_T\phi_T+A_D\phi_D$, with both amplitudes nonzero,
or perform separate pure-mode relaxations. Suppose the observation
returns the labeled material-displacement
projections
\begin{equation}
 a_i(t)=\frac{\int_{S_a}u(t)\cdot\phi_i\,dA}
                    {\int_{S_a}\|\phi_i\|^2\,dA},
 \qquad i=T,D .
 \label{eq:observe}
\end{equation}
For nonzero initial amplitudes and distinct fixed times $t_1,t_2$,
the noiseless rate summary
\begin{equation}
 r_i=\frac{\log|a_i(t_1)/a_i(t_2)|}{t_2-t_1}
 \label{eq:rateobs}
\end{equation}
eliminates the unknown amplitude exactly. Thus, with $G_s,C_a$
calibrated, the parameter-to-rate map is globally invertible:
\begin{align}
 \mu_s&=G_s/r_T,& \mu_s'&=G_s(1+2C_a)/r_D,
 \label{eq:viscosityinverse}\\
 \frac{\partial(\log r_T,\log r_D)}
      {\partial(\log\mu_s,\log\mu_s')}&=-I_2 .
 \label{eq:modalinverse}
\end{align}
The condition number is one in these log-rate coordinates with
equal row weights. Raw time samples and fitted noisy rates have
their own sensitivity and weighting, which are assessed in the
specified observation coordinates.

For this toroidal/radial pair, the normal observable annihilates
$\phi_T$ and resolves only $r_D$. Material tangential displacement
supplies the other labeled component in Eq.~(\ref{eq:observe}).
The poloidal release in Sec.~\ref{sec:releases} instead makes
both rates visible in shape at distinct rates. Excitation and observation must
therefore be specified together.

Calibration changes the result even when both rates are resolved.
If $G_s$ is co-fitted, the transformation
\begin{equation}
 (G_s,\mu_s,\mu_s')\mapsto
 (\xi G_s,\xi\mu_s,\xi\mu_s'),\qquad \xi>0,
 \label{eq:scalesym}
\end{equation}
preserves all these relaxation records. The rates then determine
$\mu_s'/\mu_s=(1+2C_a)r_T/r_D$ for known $C_a$.
If $C_a$ is also free, $r_D$ determines only
$(1+2C_a)/\mu_s'$ after fixing $G_s$.
Additional angular degrees $\ell$ repeat the same two rates in
this membrane-only model and cannot remove either calibration ambiguity.
Bulk coupling in Sec.~\ref{sec:fluid} introduces degree-dependent rates.

\subsection{Sensitivity with fitted amplitudes}\label{sec:sensitivity}
A change in a relaxation rate changes the recorded signal, but part
of that change may also be reproduced by adjusting a fitted amplitude.
We measure the rate sensitivity that remains after allowing those
amplitude adjustments. For a stacked observation vector, let $J$
contain derivatives with respect to the stated material or rate
coordinates, and let $N$ contain derivatives with respect to the
fitted nuisance coordinates, including the unknown amplitudes.

For a fixed positive-definite weight $W=L^{\mathsf T}L$, multiply
both derivative blocks by $L$ and project away the span of $LN$:
\begin{equation}
 M_W=P_{LN}^\perp LJ .
 \label{eq:weighted}
\end{equation}
Here $P_{LN}^{\perp}=I-(LN)(LN)^\dagger$ is the orthogonal
projector onto the complement of that span, and $\dagger$ denotes
the Moore--Penrose inverse. For a parameter direction $\delta\theta$, the
remaining signal change has the equivalent characterization
\begin{equation}
 \|M_W\delta\theta\|^2
 =\min_b (J\delta\theta+Nb)^{\mathsf T}W(J\delta\theta+Nb).
 \label{eq:residual}
\end{equation}
Minimization over the nuisance increment $b$ removes the part
that a fitted-amplitude change can explain. The smallest singular
value of $M_W$ measures the weakest remaining response to a unit
parameter change in the chosen coordinates. A positive value
establishes local, first-order separation of those parameters from
the fitted nuisance changes \citep{Jacquez1985,Eisenberg2014}.
The spectral condition number is the ratio of the largest to the
smallest singular value when the matrix has full column rank.

For amplitudes that enter linearly, this construction uses the
projection underlying variable projection, which eliminates linear
coefficients from a nonlinear least-squares fit
\citep{GolubPereyra1973,GolubPereyra2003}. The supplementary material
gives the general rank identities and bounds on the effect of
weighting.

\section{Paired normal records with relative-amplitude calibration}
\label{sec:paired}
\subsection{A fixed normal-only separation}\label{sec:separation}
Perform two separate releases on the same material and harmonic,
with initial displacements $A\nabla_\Omega Y_{\ell m}$ and
$C\phi_D$. Equations~(\ref{eq:poloidal}) and (\ref{eq:rates})
give their normal harmonic records:
\begin{align}
 y_1(t)&=B_p\left(e^{-r_Tt}-e^{-r_Dt}\right),
 &B_p&=\Lambda A/2,\nonumber\\
 y_2(t)&=C e^{-r_Dt},
 &\kappa&=B_p/C\ne0.
 \label{eq:pairedrecords}
\end{align}
Assume that the signed relative amplitude $\kappa$ is supplied,
while the common absolute amplitude $C\ne0$ remains unknown.
The records share a time origin and calibrated relative observation
gain. A common unknown gain can be absorbed into $C$ for rate recovery;
the energy comparison uses calibrated displacement units. An unknown
relative gain instead becomes part of the calibration of $\kappa$.
For the ordering $r_T=m-d$, $r_D=m+d$, define
$m=(r_T+r_D)/2$ and $d=|r_T-r_D|/2$. The signed amplitude $B$ used in
Sec.~\ref{sec:collision} then equals $B_p$; exchanging the rate order
changes its sign.

For common sampling times, let $e(r)=(e^{-rt_j})_j$ and
stack the records by release. The fixed invertible transformation
\begin{align}
 T_\kappa&=
 \begin{pmatrix}\kappa^{-1}I&I\\0&I\end{pmatrix},\nonumber\\
 z=T_\kappa y&=
 \begin{pmatrix}y_1/\kappa+y_2\\y_2\end{pmatrix}
 =C\begin{pmatrix}e(r_T)\\e(r_D)\end{pmatrix}
 \label{eq:pairedtransform}
\end{align}
separates the labeled rates using normal observations only.
Two distinct times in each transformed record determine its rate
by Eq.~(\ref{eq:rateobs}), including $r_T=r_D$.
The noiseless inverse fits the unknown common amplitude together
with the two rates and requires at least two distinct common times.
The radial release supplies a reference for the dilatational
component, allowing standard linear unmixing of the prescribed
responses.

The mixing must also be applied to the errors. For a full raw
covariance $\Sigma$ and fixed positive-definite raw weight $W$,
\begin{align}
 \Omega&=T_\kappa\Sigma T_\kappa^{\mathsf T},\nonumber\\
 W_z&=T_\kappa^{-\mathsf T}WT_\kappa^{-1}.
 \label{eq:pairedweight}
\end{align}
The blocks of $\Sigma$ may contain cross-record correlations.
For example, independent equal-variance raw errors give
\begin{equation}
 \Omega=\sigma^2
 \begin{pmatrix}1+\kappa^{-2}&1\\1&1\end{pmatrix}
 \otimes I .
 \label{eq:pairedcovariance}
\end{equation}
The means are separated, but their errors are generally correlated.
A covariance-aware fit therefore treats the two records jointly.
When $W=\Sigma^{-1}$, $W_z=\Omega^{-1}$.
Transforming the mean, derivatives, nuisance column, and weight
together leaves the profiled rate Gram matrix exactly unchanged.

\subsection{Independent modal patterns}
\label{sec:independence}
Equation~(\ref{eq:pairedrecords}) uses tangential and radial
preparations. The tangential release contains both poloidal eigenmodes;
the radial release contains the dilatational eigenmode alone.
The general requirement is independence of their modal patterns.

Let two arbitrary poloidal releases $q^{(1)},q^{(2)}$ be prescribed at
the same harmonic. Decomposing each on the eigenfamilies of
Eq.~(\ref{eq:poloidal}) and reading off the normal component gives the
release-to-record map
\begin{equation}
 (P,Q)=\left(\tfrac{1}{2}\Lambda q_p,\;
              q_r-\tfrac{1}{2}\Lambda q_p\right),
 \label{eq:releasemap}
\end{equation}
so that each normal record is
\begin{equation}
 y_j(t)=S\left(P_je^{-r_Tt}+Q_je^{-r_Dt}\right),
 \qquad j=1,2,
 \label{eq:generalrecords}
\end{equation}
with $(P_j,Q_j)$ known from the prescribed releases and $S\ne0$ a
single unknown common scale. Collect
\begin{equation}
 \mathcal M=\begin{pmatrix}P_1&Q_1\\P_2&Q_2\end{pmatrix}.
 \label{eq:modalmatrix}
\end{equation}
Whenever $\det\mathcal M\ne0$, the fixed transformation $\mathcal M^{-1}$ replaces
$T_\kappa$ and yields the same two labeled records
$S\,e(r_T)$ and $S\,e(r_D)$, with the weight transformed as in
Eq.~(\ref{eq:pairedweight}). The tangential/radial pair of
Eq.~(\ref{eq:pairedrecords}) is the instance
$\mathcal M=\bigl(\begin{smallmatrix}B_p&-B_p\\0&C\end{smallmatrix}\bigr)$,
for which $\det\mathcal M=B_pC$, and $\kappa$ is the one ratio needed to
invert it.

Modal independence, $\det\mathcal M\ne0$, is therefore a preparation
criterion: any calibrated mixed releases with an invertible modal
matrix are admissible. In this reference model the eigenfamilies are
fixed, so the criterion can be evaluated before taking relaxation data.
Section~\ref{sec:couplingtradeoff} treats the parameter-dependent modal
patterns introduced by fluid coupling.

This also identifies the exact-coincidence mechanism. At $r_T=r_D=r$
the two rate columns of Eq.~(\ref{eq:generalrecords}) are
$S(P\otimes v)$ and $S(Q\otimes v)$, with $v=(-t_je^{-rt_j})_j$.
Throughout, $\otimes$ orders the release coordinate before the time coordinate. These are linearly
independent precisely when $P$ and $Q$ are not proportional across
releases, which is $\det\mathcal M\ne0$. A single release makes them
proportional identically, and that is the source of the single-record
degeneracy quantified in Sec.~\ref{sec:collision}.

At coincidence, a common schedule $t_1<\dots<t_n$, $n\ge2$, and a
factorized weight $W=W_r\otimes W_t$ give an explicit sharp bracket.
Here $W_r$ and $W_t$ are positive-definite release and time weights;
the unweighted case uses identity matrices. The more general bound in
Eq.~(\ref{eq:pairedbound}) permits any positive-definite weight.

At $r_T=r_D=r$ put $e=(e^{-rt_j})_j$ and $v=(-t_je^{-rt_j})_j$. The
two rate columns are $S(P\otimes v)$ and $S(Q\otimes v)$, where
$\otimes$ pairs a release-space vector with a time-space vector, so
the unprojected rate block $G$ has
\begin{equation}
 \begin{aligned}
  G^{\mathsf T}WG&=S^2\|v\|_{W_t}^2\,\mathcal M^{\mathsf T}W_r\mathcal M,\\
  \sigma_{\min}(LG)&=|S|\,\|v\|_{W_t}\,\sigma_{\min}(W_r^{1/2}\mathcal M),
 \end{aligned}
 \label{eq:coincidencegram}
\end{equation}
exactly, because
$(w\otimes v)^{\mathsf T}(W_r\otimes W_t)(z\otimes v)
 =(w^{\mathsf T}W_rz)(v^{\mathsf T}W_tv)$. The amplitude column is
$(P+Q)\otimes e$, which lies in the subspace paired with
$\operatorname{span}(e)$, while $\operatorname{range}(G)$ is paired
with $\operatorname{span}(v)$. Removing one direction can reduce a
vector only by the sine of its angle to that direction, and every
element of $\operatorname{range}(G)$ meets $(P+Q)\otimes e$ at an
angle at least $\angle_{W_t}(v,e)$, since
$|\langle w\otimes v,z\otimes e\rangle_W|
 \le\|w\|_{W_r}\|z\|_{W_r}\,|\langle v,e\rangle_{W_t}|$. Hence, writing
$\sigma_\star=|S|\,\|v\|_{W_t}\,\sigma_{\min}(W_r^{1/2}\mathcal M)$,
\begin{equation}
 \sin\angle_{W_t}(v,e)\,\sigma_\star
 \le\sigma_{\min}(M_\kappa)
 \le\sigma_\star,
 \label{eq:coincidencebracket}
\end{equation}
with both outer factors depending only on the schedule, the time weight
$W_t$, the coincident rate $r$ and the amplitude scale. Because the two
singular values of $W_r^{1/2}\mathcal M$ multiply to
$\det(W_r)^{1/2}|\det\mathcal M|$, both ends of the bracket are
proportional to $|\det\mathcal M|$ at fixed
$\sigma_{\max}(W_r^{1/2}\mathcal M)$. The coincidence margin therefore
vanishes exactly when $\det\mathcal M=0$ and lies between two multiples of
$|\det\mathcal M|$ whose ratio is at most $1/\sin\angle_{W_t}(v,e)$. Its position within the bracket depends on the singular values of
$W_r^{1/2}\mathcal M$ and on how the amplitude direction
$W_r^{1/2}(P+Q)$ meets their singular directions. The Supplementary
Material, Sec.~S1, constructs two modal matrices with identical singular
values at opposite ends. The quantity $\sigma_{\min}(W_r^{1/2}\mathcal M)$
fixes the bracket, while the amplitude direction sets the margin within it.

The endpoint constructions in Supplementary Sec.~S1\,A establish
sharpness. For a general positive-definite weight, the eigenvalue
sandwich in Supplementary Eq.~(S3) controls the margin instead.
The supplementary observation designs test both cases and the common-
schedule assumption. Together, these results separate the geometric
benefit of independent releases from the effects of sampling and noise
weighting.

\subsection{Sensitivity through rate coincidence}
Use labeled linear rate coordinates $(r_T,r_D)$ in fixed units.
Let $J_r=\partial y/\partial(r_T,r_D)$ and
$n_C=\partial y/\partial C$, and choose $L^{\mathsf T}L=W$.
The weighted rate sensitivity with $C$ co-fitted is
$M_\kappa=P_{Ln_C}^{\perp}LJ_r$, as in
Eq.~(\ref{eq:weighted}). Fix at least two distinct nonnegative times,
write $t_{\max}=\max_jt_j$, and set
\begin{align}
 S_t&=\sum_j(t_j-\bar t)^2>0,\nonumber\\
 \bar t&=\frac{1}{n}\sum_jt_j .
\end{align}
For fixed nonzero $\kappa$, fixed SPD $W$, and rates in a compact
positive interval $[r_{\min},r_{\max}]$, the explicit bound is
\begin{equation}
 \sigma_{\min}(M_\kappa)\ge
 |C|\sqrt{\lambda_{\min}(W_z)S_t}\,
 e^{-r_{\max}t_{\max}} .
 \label{eq:pairedbound}
\end{equation}
The two local rate changes remain distinguishable as their rates
coincide. Uniformity over an amplitude class requires
$|C|\ge C_{\min}>0$. For isotropic raw covariance $\sigma^2 I$,
\begin{equation}
 \lambda_{\min}(W_z)=
 \frac{1+2\kappa^2-\sqrt{1+4\kappa^4}}{2\sigma^2}.
 \label{eq:kappaboundweight}
\end{equation}
It behaves as $\kappa^2/\sigma^2$ for small $|\kappa|$ and tends
to $1/(2\sigma^2)$ for large $|\kappa|$. At fixed total energy,
$|C|$ also decreases for large $|\kappa|$ by
Eq.~(\ref{eq:pairedamplitude}). Thus gap uniformity applies to the specified release and sampling
design.
At the coincident-rate design of Fig.~\ref{fig:paired}, the actual
smallest rate singular value is about $15.2$ times this conservative
bound, using $r_{\max}=m$ for that reference calculation.

To prove the bound, write $v_i=-t e(r_i)$.
The transformed rate columns are $C(v_T;0)$ and $C(0;v_D)$,
and the amplitude column is $(e(r_T);e(r_D))$.
For a unit rate direction $q$, apply the residual formula
(\ref{eq:residual}), bound $W_z$ below by its smallest eigenvalue,
and relax the common amplitude coefficient to two independent
channel coefficients. The result is
\begin{align}
 \|M_\kappa q\|^2
 &\ge C^2\lambda_{\min}(W_z)
       \sum_{i=T,D}q_i^2\mathcal V(r_i),\nonumber\\
 \mathcal V(r)
 &=\min_b\sum_j e^{-2rt_j}(t_j-b)^2\nonumber\\
 &\ge e^{-2r_{\max}t_{\max}}S_t .
 \label{eq:pairedvariance}
\end{align}
This gives Eq.~(\ref{eq:pairedbound}) directly at rate coincidence,
without a singular nuisance limit.
The full Jacobian in $(r_T,r_D,C)$ has rank three because
$e(r)$ and $t e(r)$ are independent at distinct times.
The full-column-rank Jacobian therefore gives a locally injective
noiseless map in these coordinates.

With calibrated elasticity, the log-viscosity Jacobian is
$-J_r\diag(r_T,r_D)$, so a corresponding lower bound has the
additional factor $r_{\min}$.
For $r_T=m-d,r_D=m+d$, the change to $(m,d)$ coordinates
multiplies both singular values by $\sqrt{2}$.
These quantitative bounds measure local sensitivity in the specified
observation and parameter coordinates.

\subsection{Finite initial energy and release-pair selection}
For this energy comparison, $A,B_p,C$ denote physical length
amplitudes in calibrated displacement units, without an
absorbed unknown observation gain.
Set $c=1+2C_a$ and $\chi=c+1-2/\Lambda>0$.
The energy $K(u,u)/2$ from Eqs.~(\ref{eq:formshear}), (\ref{eq:forms}) and
(\ref{eq:stiffness}) give, for the two separate initial states,
\begin{align}
 E_{\rm tan}
 &=\frac{G_sA^2}{2}
       [\Lambda(\Lambda-2)+c\Lambda^2]\nonumber\\
 &=2G_s\chi B_p^2,\qquad
 E_{\rm rad}=2G_scC^2 .
 \label{eq:pairedenergy}
\end{align}
At fixed positive total comparison energy $E_0$,
\begin{equation}
 |C|=\sqrt{\frac{E_0}{2G_s(\chi\kappa^2+c)}},
 \qquad B_p=\kappa C .
 \label{eq:pairedamplitude}
\end{equation}
For fixed supplied $\kappa\ne0$, neither excitation grows as the
rate gap closes. At fixed harmonic degree and geometry, sufficiently
small $E_0$ keeps the displacement gradients small. For fixed $E_0>0$
and elastic parameters in a compact positive set,
$2G_s(\chi\kappa^2+c)$ is bounded above and away from zero, so
Eq.~(\ref{eq:pairedamplitude}) supplies a nonzero amplitude lower bound.
The two-release protocol therefore retains
Eq.~(\ref{eq:pairedbound}) without an inverse-gap strain cost.
By contrast, keeping the single tangential normal signal coefficient
$\hat Q=2Bd$ fixed requires $|B|\propto1/d$, as shown in
Sec.~\ref{sec:collision}.

This comparison fixes the energy of the prepared states while fitting
$C$ in the inverse. It therefore isolates the benefit of relative
calibration without supplying an absolute-amplitude measurement.
An upper energy budget alone permits arbitrarily weak excitation;
the positive bound uses the fixed nonzero energy in
Eq.~(\ref{eq:pairedamplitude}).

The release pair is also a design variable. At $\ell=2$, natural
candidates are pure radial and tangential displacements, the trace-free
mode $q_r/q_p=\Lambda/2$, and the mode
$q_r/q_p=[\Lambda(\Lambda-2)+c\Lambda^2]/(2c\Lambda)$.
The last two require purely tangential and purely normal holding
tractions, respectively. Comparing their sensitivities requires a
common budget because the margin is homogeneous of degree one in a common release amplitude.
Define the spread as the largest divided by the smallest projected
sensitivity across the six admissible pairs. For the reference
coincident-rate design at $\ell=2$ and $C_a=2$, with four unweighted
sampling times specified in Sec.~S4.B of the supplementary material,
the spread is $9.24$ under equal coordinate amplitude,
$q/\|q\|_2$, and $3.88$ under equal initial energy. The trace-free and normal-traction-only
pair is the least sensitive there: its holding tractions are purely
tangential and purely normal, respectively, but its modal patterns
are nearly parallel. Thus preparation simplicity and sensitivity
need not favor the same pair. The supplementary
material gives the specified algebraic design survey and distinguishes
its positive-stiffness controls from the physical area-modulus range.

\subsection{Sensitivity with independent unknown amplitudes}\label{sec:uncalibrated}
With independent unknown amplitudes $B_p$ and $C$, the radial
reference improves positive-gap sensitivity, which tends to zero
linearly at coincidence.
Let $\Delta=r_T-r_D$ be the full signed gap and hold $r_D=r$.
Put $e=e(r)$, $v=-te$, and $w=t^2e$, and define
\begin{align}
 \mathcal U&=\operatorname{span}
       \{(v;0),(0;e),(0;v)\},\nonumber\\
 D_W(r)&=\operatorname{dist}_W((w;0),\mathcal U)>0 .
\end{align}
Here distance is measured in the raw $W$ norm.
For fixed nonzero $B_p,C$, fixed SPD $W$, and at least two
distinct positive times, the rate sensitivity projected over
both independent amplitudes satisfies
\begin{equation}
 \sigma_{\min}(M_{\rm ind})
 =\frac{|B_p|}{2}D_W(r)|\Delta|+O(\Delta^2).
 \label{eq:pairedfreegap}
\end{equation}
The other singular value has a positive limit. Thus a radial
reference improves the observation, but relative calibration
is what retains sensitivity as the rates approach each other.
Positivity of $D_W$ follows because $w$ is not proportional
to $v$ at distinct positive times.

For the coefficient, set $h_\Delta=e(r+\Delta)-e(r)$.
For $\Delta\ne0$, rescale the first nuisance column to
$(h_\Delta/\Delta;0)$ without changing its span.
Together with $(0;e)$, it extends smoothly to the independent
columns $(v;0),(0;e)$ at zero.
If $\Pi_\Delta$ projects off their whitened span, subtracting
the rescaled nuisance column from the first rate column gives
\begin{align}
 &\Pi_\Delta L\begin{pmatrix}B_p e'(r+\Delta)\\0\end{pmatrix}
 \nonumber\\
 &\quad=\frac{B_p\Delta}{2}\Pi_0L
       \begin{pmatrix}w\\0\end{pmatrix}
       +O(\Delta^2).
\end{align}
The second projected rate column tends to
$C\Pi_0L(0;v)$.
Eliminating that nonzero limiting column gives
Eq.~(\ref{eq:pairedfreegap}).

At exact coincidence the actual $B_p$-nuisance derivative is
zero, so the actual nuisance span drops to $(0;e)$.
Projection of $(B_pv;0)$ and $(-B_pv;Cv)$ over that smaller
span gives rank two. This singular nuisance point has a different projection from the
nonzero-gap limit, whose sensitivity tends to zero. The noiseless rates are still fixed:
for $B_p\ne0$, $y_1(t)\equiv0$ forces $r_T=r_D$, and the
radial record fixes their common value.
The numerical treatment of this small nuisance column is described
in Sec.~\ref{sec:verification}.

\subsection{Finite calibration information}
\begin{figure*}[t]
 \includegraphics[width=\textwidth]{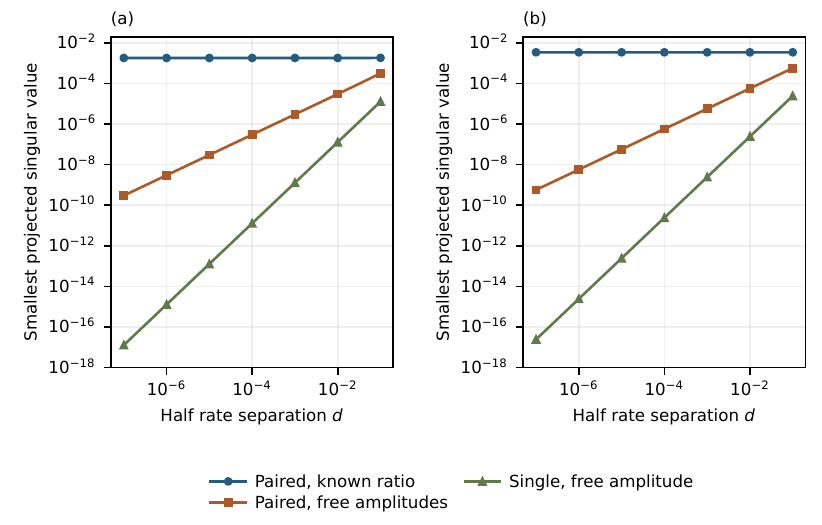}
 \caption{Covariance-aware sensitivity versus the half-gap
 $d=|r_D-r_T|/2$: (a) linear labeled-rate coordinates and
 (b) calibrated log-viscosity coordinates.
 The paired normal protocol with known relative amplitude,
 the same pair with independent unknown amplitudes, and a
 single tangentially excited record with unknown amplitude
 are compared at the same total initial energy.
 The fixed mean rate is $m=(r_T+r_D)/2=1.9$, with $\kappa=1.3$,
 $G_s=1$, $C_a=2$, $\ell=2$, $E_0=0.01$, and
 $t_j=(0.05,0.11,0.23,0.42,0.7,1.05,1.6,2.4)$,
 in the reference units of Sec.~\ref{sec:verification}.
 The paired calculation uses the prescribed correlated covariance
 $\Sigma_{\rm X}$ of the Supplementary Material, Eq.~(S41) in Sec.~S4\,A,
 in units $u_0^2$ with $W=\Sigma^{-1}$; the single record uses its
 first-record marginal. $\Sigma_{\rm X}$ combines autoregressive time correlation, unequal
record variances and cross-record correlation with a fixed rank-one term.
The full covariance is retained under mixing, and the ordinates scale
inversely with a common noise amplitude.
 The energy constraint fixes the total preparation budget; the paired
 protocols use two releases and 16 samples, the single record one release
and 8 samples. The comparison concerns small-gap orders at equal energy;
the finite-gap levels also depend on acquisition count and covariance.}
 \label{fig:paired}
\end{figure*}

Fixed positive local information about $\kappa$ suffices for a
local uniform-information result. For a small fixed calibration
error, suppose the fit uses $\kappa+\delta\kappa$ while the true record
uses $\kappa$. With the raw weight fixed and $C$ profiled, define
$g=\partial_\kappa y$ and $b=P_{Ln_C}^{\perp}Lg$.
At a full-rank reference, the first-order weighted least-squares shift is
\begin{equation}
 \delta r=-M_\kappa^\dagger b\,\delta\kappa+O(\delta\kappa^2).
 \label{eq:kappabias}
\end{equation}
The coefficient depends on the specified rates, schedule, covariance
and release amplitude. At exact coincidence
$g=0$, so this scalar-ratio error has no first-order rate bias.
The release-composition error in Eq.~(\ref{eq:areaerror}) enters the
joint preparation map described above.
As an explicit assumed error model, suppose the raw
errors are Gaussian with covariance $\Sigma$ and an independent
calibration datum for $\kappa$ has fixed positive information
$\mathcal I_\kappa$. For example, a Gaussian calibration datum
with known variance $\tau^2$ gives $\mathcal I_\kappa=\tau^{-2}$.
Fit $\kappa$ and $C$ jointly, using $W=\Sigma^{-1}$.
Here $\kappa$ includes the relative observation gain; the common
time origin remains supplied as in Sec.~\ref{sec:separation}.
Let $h=e(r_T)-e(r_D)$, $g=\partial y/\partial\kappa=C(h;0)$,
and, after profiling $C$, put
\begin{align}
 F_\kappa&=M_\kappa^{\mathsf T}M_\kappa,&
 b&=P_{Ln_C}^{\perp}Lg,\nonumber\\
 u&=M_\kappa^{\mathsf T}b,& v_\kappa&=\|b\|^2 .
\end{align}
The remaining Schur complement is
\begin{align}
 F_{\rm cal}
 &=F_\kappa-\frac{uu^{\mathsf T}}
                      {v_\kappa+\mathcal I_\kappa}\nonumber\\
 &\succeq
 \frac{\mathcal I_\kappa}{v_\kappa+\mathcal I_\kappa}
 F_\kappa .
 \label{eq:pairedcalibration}
\end{align}
The inequality follows from
$uu^{\mathsf T}\preceq v_\kappa F_\kappa$.
The information lower factor is at least one half if
$\mathcal I_\kappa\ge v_\kappa$, or $\tau\le v_\kappa^{-1/2}$
for the stated Gaussian calibration datum. In the units, energy and
covariance of Fig.~\ref{fig:paired}, $v_\kappa\simeq9.75\times10^{-7}$
at half-gap $d=0.1$ for that numerical covariance and preparation energy.
On a fixed compact design, $v_\kappa$ is bounded, so any fixed
$\mathcal I_\kappa>0$ retains a positive uniform lower bound.
Moreover $g=O(\Delta)$ implies
$F_{\rm cal}=F_\kappa-O(\Delta^2)$ at rate coincidence.
The zero-calibration-information and zero-gap limits do not
commute. Removing calibration information at nonzero gap recovers
the independent-amplitude projection, whose weak singular value tends
to zero linearly. Taking the gap to zero with fixed positive information
instead gives the positive fixed-ratio Gram matrix $F_\kappa$;
its subsequent zero-information limit retains that matrix.

Thus an uncertain ratio is co-fitted with an informative calibration
datum. A hard interval alone supplies zero local likelihood information
at an interior point.
Figure~\ref{fig:paired} compares the finite-amplitude paired
sensitivities with these calibration distinctions.

\section{Finite-time recovery of the constrained shape response}
\label{sec:collision}
\subsection{An ordered inverse from three positive times}
Order the distinct rates as $m-d$ and $m+d$, with $m>d>0$.
Absorb their possible exchange into the signed amplitude $B\ne0$,
so Eq.~(\ref{eq:poloidal}) becomes
\begin{equation}
 y(t)=2B e^{-mt}\sinh(dt).
 \label{eq:gapresponse}
\end{equation}
All times and rate coordinates in this section use fixed reference
units. For any $0<t_1<t_2<t_3$, noiseless model data have the
same nonzero sign. Let $l_i=\log|y(t_i)|$ and define
\begin{align}
 \Gamma_{\rm obs}&=\frac{l_3-l_2}{t_3-t_2}
                  -\frac{l_2-l_1}{t_2-t_1},\\
 F(d)&=\frac{\log\sinh(dt_3)-\log\sinh(dt_2)}{t_3-t_2}
       \nonumber\\
 &\quad-\frac{\log\sinh(dt_2)-\log\sinh(dt_1)}{t_2-t_1}.
 \label{eq:threeinverse}
\end{align}
The amplitude and mean rate cancel, leaving $F(d)=\Gamma_{\rm obs}$.
This scalar equation has exactly one positive solution for model
data. To see this, put $g_d(t)=t\coth(dt)$; then
\begin{equation}
 g_d''(t)=2d\,\operatorname{csch}^2(dt)
                    [dt\coth(dt)-1]>0 .
\end{equation}
Thus $F'(d)>0$, because it is the difference of successive secant
slopes of a strictly convex function. Moreover
\begin{align}
 \Gamma_0&=\frac{\log(t_3/t_2)}{t_3-t_2}
       -\frac{\log(t_2/t_1)}{t_2-t_1}<0,\\
 F(0+)&=\Gamma_0,\qquad F(\infty)=0 .
\end{align}
After recovering $d$, set
\begin{align}
 m&=\frac{\log[\sinh(dt_2)/\sinh(dt_1)]-(l_2-l_1)}
               {t_2-t_1},\label{eq:recoverm}\\
 B&=\frac{y(t_1)e^{mt_1}}{2\sinh(dt_1)}.
 \label{eq:recovermb}
\end{align}
Common sign, $\Gamma_0<\Gamma_{\rm obs}<0$, and $m>d$ are necessary
and sufficient for three nonzero data values to belong to this
positive-rate model, with no additional amplitude bound imposed.
An imposed small-displacement bound must also be checked on $B$.
For two positive times, every nearby $d$ can instead be paired
with $m$ and $B$ from the recovery formulas; hence three
is the minimum number of scalar samples on the open parameter
domain. The known zero at $t=0$ adds no constraint to this noiseless
inverse. With independent errors it also adds no rate information;
with correlated errors it can reduce the conditional noise at other times.
The ordered rates are unique; assigning them to shear or dilation
still needs the signed excitation or another mode label.

Uniqueness also follows from the classical real-exponential
Chebyshev property \citep{Aldaz2009}: the difference of two
candidate signals contains at most four distinct exponentials and
vanishes at the three positive times and at zero. A nonzero such
exponential polynomial has at most three zeros. The scalar
construction above additionally supplies the inverse and exact
data-membership conditions.
The known zero at the origin makes three positive samples sufficient
for the constrained response, compared with four samples for a
general two-exponential response with two independent amplitudes.
These logarithmic formulas give the noiseless inverse; fitting
additive-noise data requires the observation error model. In particular,
\begin{equation}
 F(d)=\Gamma_0+\frac{t_3-t_1}{6}d^2+O(d^4),
\end{equation}
so the inverse becomes sensitive as the two rates approach.

\subsection{Fixed-amplitude conditioning near coincident rates}
Let $t_i$ be a fixed finite set containing at least three distinct
positive times, and let $W=L^{\mathsf T}L$ be fixed and positive
definite. Keep $m>0$ and $B\ne0$ fixed while $d\to0+$.
Write $x_k=L(t_i^k e^{-mt_i})_i$ and define
\begin{equation}
 a_2=P_{x_1}^{\perp}x_2,\qquad
 b_3=P_{[x_1,x_2]}^{\perp}x_3 .
\end{equation}
Both vectors are nonzero: restriction to three distinct positive-time
rows before whitening gives a nonzero diagonal factor times
$(1,t_i,t_i^2)$. Invertible whitening preserves column independence.
Let $J=\partial y/\partial(m,d)$ and
$N=\partial y/\partial B$. The two singular values of
$M_W=P_{LN}^{\perp}LJ$ satisfy
\begin{align}
 \sigma_{\max}(M_W)&=2|B|\,\|a_2\|d+O(d^3),\\
 \sigma_{\min}(M_W)&=\frac{2|B|}{3}\,\|b_3\|d^2+O(d^4).
 \label{eq:collisionlaw}
\end{align}
Consequently
\begin{equation}
 \operatorname{cond}(M_W)\sim\frac{3\|a_2\|}{\|b_3\|d},\qquad
 \lambda_{\min}(M_W^{\mathsf T}M_W)
       \sim\frac{4B^2}{9}\|b_3\|^2d^4 .
 \label{eq:collisioninfo}
\end{equation}
These are raw-sample rate sensitivities, distinct from the
log-rate summary in Eq.~(\ref{eq:modalinverse}).
Changing to the two log-viscosity coordinates at calibrated
$G_s,C_a$ changes the constants but preserves the powers:
the coordinate transformation and its inverse stay bounded
as $d\to0$ with $m>0$.

For completeness, expand the whitened signal as
$Ly=2B(dx_1+d^3x_3/6+O(d^5))$.
The nuisance direction tends to $x_1$. Eliminating it gives
\begin{align}
 (M_W)_m&=-2Bd\,a_2+O(d^3),\\
 (M_W)_d&=\frac{2B}{3}d^2 P_{x_1}^{\perp}x_3+O(d^4).
\end{align}
The second expression follows by subtracting $(B/d)LN$
from $L\,\partial_d y$ before projection. Removing the
component parallel to the first column gives
Eq.~(\ref{eq:collisionlaw}). The constants encode the fixed observation window and correlation
weights; windows or weights that vary with $d$ require a separate
uniformity analysis.

At $d=0$, $N=0$, so the projector is the identity and the
pointwise rate Jacobian is $[0,2Bx_1]$, of rank one.
For $d>0$ the projected matrix has rank two but tends to zero.
The nuisance rank changes at coincidence, making its projector
discontinuous. The exact zero-signal family at $d=0$ must be
analyzed directly, rather than by extrapolating the regular
local-inverse criterion.

\subsection{Calibration, signal normalization, and finite errors}
If the signed amplitude is known, no amplitude projection is
needed. The same expansion gives
\begin{align}
 \sigma_{\max}(LJ)&=2|B|\|x_1\|+O(d^2),\\
 \sigma_{\min}(LJ)&=2|B|\|a_2\|d+O(d^3).
 \label{eq:knownamplitude}
\end{align}
Thus amplitude calibration changes the weak-sensitivity power.
Both cases nevertheless have condition number $O(d^{-1})$:
the condition number alone hides their different absolute
sensitivity losses.
It cannot remove the exact common-rate ambiguity at $d=0$.
In contrast, separately labeled nonzero modal records, each
with at least two distinct sample times, retain two nonzero
amplitude-projected rate sensitivities there. This comparison assesses rank for the specified observables; matching
sensor cost or total elastic energy requires a separate acquisition
design.

Holding $\hat Q=2Bd$ fixed as $d\to0$ would make
$y=\hat Q e^{-mt}\sinh(dt)/d$ approach a finite signal.
In the original excitation this requires $|B|\propto1/d$.
At fixed harmonic degree, $B=\Lambda A/2$ and the initial
strain scale is proportional to $|A|/a$; such a sequence
eventually violates any fixed small-strain bound.
Rescaling both data and noise consistently is only a coordinate
change. Renormalizing the signal while retaining an unchanged
absolute-noise level describes a different problem.

This distinction also specifies the relation to the Bateman response
\citep{Garrett1994}. For first-order transfer and removal, its prefactor
contains the transfer rate divided by the rate difference. Holding
the underlying dose and volume factors fixed gives a finite
$t e^{-rt}$ limit at equal rates. Holding the membrane displacement
$B$ fixed instead gives the zero normal response in
Eq.~(\ref{eq:gapresponse}). Identical exponential functions therefore
need not have identical admissible coincidence paths or inverse problems.

\begin{figure*}[t]
 \includegraphics[width=\textwidth]{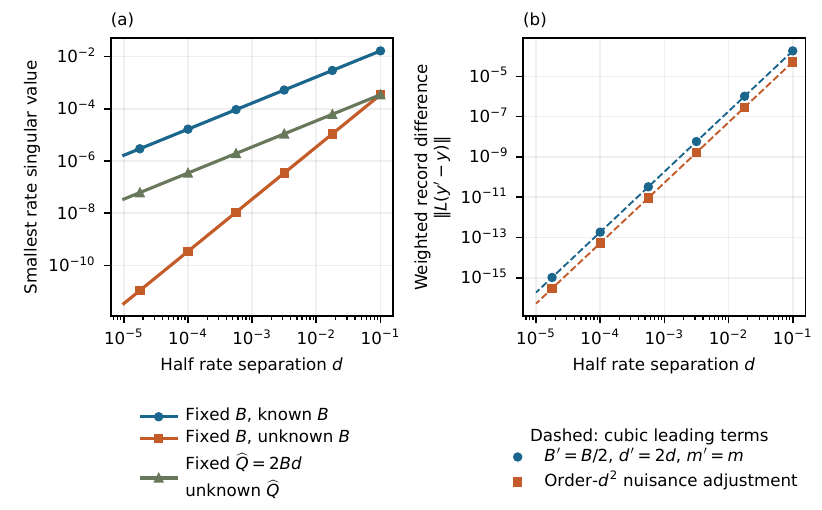}
 \caption{Finite-time conditioning of the signed normal response.
 (a) Smallest rate-sensitivity singular value with unknown amplitude,
 known amplitude, and a fixed leading signal coefficient $\hat Q=2Bd$.
 The last of these
 requires excitation growing inversely with the half-gap and is
 excluded from a fixed small-strain comparison.
 (b) Weighted record differences $\|L(y'-y)\|$ for a doubled-half-gap alternative with
 reduced excitation, before and after the asymptotic amplitude
 and mean-rate adjustments. Dashed asymptotes use the analytic
 prefactors, rather than fitted slopes. The reference values are
 $m=1.9$, $B=0.6$, $W=I$, and
 $t_i=(0.05,0.11,0.23,0.42,0.7,1.05,1.6,2.4)$;
 the rescaled curve holds $\hat Q=0.12$ fixed. Rates, time, and
 amplitude use the fixed reference units of
 Sec.~\ref{sec:verification}. At fixed rates, the unscaled-record
 sensitivities and record differences scale linearly with $|B|$.
 Both alternatives in (b) give cubic differences; the nuisance
 adjustment changes their leading coefficient.}
 \label{fig:collision}
\end{figure*}

A finite alternative makes the loss of resolution explicit
without choosing a stochastic noise law or fitting algorithm.
Fix $k>1$ and compare $(B,m,d)$ with $(B/k,m,kd)$.
Both amplitudes remain within the same bound if the first does,
and both rates are positive for sufficiently small $d$. Then
\begin{equation}
 L[y(B/k,m,kd)-y(B,m,d)]
     =\frac{B(k^2-1)}{3}d^3x_3+O(d^5).
 \label{eq:finitealternative}
\end{equation}
Small $O(d^2)$ adjustments to the alternative amplitude and
mean rate can remove the $x_1,x_2$ components; the resulting
leading distance is
\begin{equation}
 \frac{|B|(k^2-1)}{3}\|b_3\|d^3+O(d^5).
 \label{eq:finitedistance}
\end{equation}
The supplementary material constructs those adjustments.
The half-gaps differ by $(k-1)d$, a fixed relative amount,
although their data distance is cubic. If the weighted
absolute-error radius is at least half that distance, the
midpoint data are compatible with both models. This gives a deterministic obstruction to uniform gap resolution
at that absolute-error scale.
This is the classical difficulty of distinguishing nearby exponential
decay rates, as studied through the geometry of Prony reconstruction with near-colliding nodes \citep{Akinshin2021}. In node coordinates $x_\pm=x_c\pm h$
with amplitudes $(B,-B)$, the replacement $(B,h)\mapsto(B/k,kh)$
preserves moments of orders zero, one, and two, while changing
the third by $2B(k^2-1)h^3$. It lies on the classical two-node
Prony curve; the finite-time formulas above specify its weighted
membrane-record manifestation. Figure~\ref{fig:collision} tests these
fixed-amplitude laws numerically.

\subsection{A matrix-defined comparison}
Table~\ref{tab:comparison} separates rate sensitivity after fitting
amplitudes from sensitivity to all fitted parameters together.
Let $\mathcal J=L[N,J]$ be the full joint Jacobian and
$M=P_{LN}^{\perp}LJ$ the amplitude-projected rate block.
All powers are for $d\to0+$ at fixed nonzero amplitudes, fixed
distinct positive times, fixed SPD weights, and fixed parameter units.
For the single-record rows use $(m,d)$ rate coordinates and the
listed linear amplitude coordinates. Bounded nonsingular coordinate
changes preserve these powers; singular gap-dependent changes need
not. The paired rows use labeled rates. Here the full gap is
$|\Delta|=2d$, so these powers are also the powers in $|\Delta|$.

\begin{table*}[t]
\caption{Smallest singular-value orders for specified observation
and amplitude models. The sample count is the minimum for regular
local recovery at distinct rates using distinct positive times;
paired counts refer to the same times in each of two records.
$\Theta(1)$ denotes a finite positive
limit under the stated nonzero-amplitude assumptions.
Equal exponents describe orders, not equal constants or acquisition costs.
The two known-pattern rows assume $PQ\ne0$ and differ only in whether
the initial normal signal $S(P+Q)$ vanishes. This distinction sets
the coincidence order and whether the stated uniform global
three-positive-sample guarantee applies. It does not imply
nonuniqueness for every generic-pattern dataset. The supplementary
material gives the expansions and their sampling conditions.}
\label{tab:comparison}
\begin{ruledtabular}
\begin{tabular}{lccc}
Response and fitted amplitudes & Samples & $\sigma_{\min}(M)$ &
 $\sigma_{\min}(\mathcal J)$\\
\hline
$A e^{-(m-d)t}+C e^{-(m+d)t}$; fit $A,C$ & 4 & $\Theta(d^2)$ & $\Theta(d^3)$\\
$S[Pe^{-(m+d)t}+Qe^{-(m-d)t}]$, $P+Q\ne0$; fit $S$ & 3 &
 $\Theta(d)$ & $\Theta(d)$\\
$2B e^{-mt}\sinh(dt)$; fit $B$ & 3 & $\Theta(d^2)$ & $\Theta(d^3)$\\
$2B e^{-mt}\sinh(dt)$; supplied $B$ & 2 & $\Theta(d)$ & $\Theta(d)$\\
Paired normal records; fit $B_p,C$ & $2+2$ & $\Theta(d)$ & $\Theta(d^2)$\\
Paired normal records; supplied $\kappa$, fit $C$ & $2+2$ &
 $\Theta(1)$ & $\Theta(1)$
\end{tabular}
\end{ruledtabular}
\end{table*}

The supplied-ratio pair retains a positive limiting sensitivity,
whereas the other five models in Table~\ref{tab:comparison} lose
sensitivity as the gap closes. For a single record, the initial normal signal explains the contrast
between the two known-pattern rows. A generic pattern with $P+Q\ne0$
loses sensitivity linearly, whereas the equal-and-opposite pattern
starts at zero and loses it quadratically. That zero initial value
also supplies the extra constraint behind the three-positive-sample
inverse of Sec.~\ref{sec:collision}. For generic patterns, three
samples suffice for regular local recovery but global uniqueness is
pattern- and data-dependent. These are fixed-pattern asymptotics:
their remainder bounds need not be uniform as $P+Q\to0$. The
supplementary material gives the expansions, constants and finite-root
examples.

Classical Prony error exponents concern specified data and small-error
regimes \citep{Akinshin2021,Katz2024}. Here the two columns of
Table~\ref{tab:comparison} distinguish rate sensitivity after fitting
amplitudes from sensitivity to all fitted parameters together.

\section{A controlled spherical bulk-dissipation extension}
\label{sec:fluid}
\subsection{Resistance of the inner and outer Stokes fluids}
Keep the reference geometry, finite area elasticity, and zero-bending
stiffness, and surround the membrane by incompressible Newtonian
fluids of viscosities $\eta_{\rm in},\eta_{\rm out}>0$.
The interior flow is regular at the origin, the exterior flow decays
at infinity, and no slip or permeation occurs at $S_a$.
Spherical-harmonic solutions of the interior and exterior Stokes
problems are used in shell hydrodynamics
\citep{Kuriabova2008,KuriabovaErratum}. Quasistatic Stokes flow adds
a symmetric positive resistance
$R_f$:
\begin{equation}
 (V+R_f)\dot q+Kq=0 .
 \label{eq:fluidode}
\end{equation}
For one poloidal harmonic write
$u=q_rYn+q_p\nabla_\Omega Y$, and let $q=(q_r,q_p)^{\mathsf T}$.
In these physical length coordinates,
\begin{align}
 R_s&=\begin{pmatrix}0&0\\0&\Lambda(\Lambda-2)\end{pmatrix},\nonumber\\
 R_d&=\begin{pmatrix}4&-2\Lambda\\-2\Lambda&\Lambda^2\end{pmatrix}.
 \label{eq:poloidalmatrices}
\end{align}
The fluid block, with dissipation $\dot q^{\mathsf T}R_f\dot q$, is
\begin{align}
 R_{f,\ell}&=a\begin{pmatrix}\rho_{rr}&\rho_{rp}\\
                               \rho_{rp}&\rho_{pp}\end{pmatrix},\nonumber\\
 \rho_{rr}
 &=\eta_{\rm in}(2\ell+1+3/\ell)\nonumber\\
 &\quad+\eta_{\rm out}[2\ell+1+3/(\ell+1)],\nonumber\\
 \rho_{rp}&=-3[\eta_{\rm in}(\ell+1)+\eta_{\rm out}\ell],\nonumber\\
 \rho_{pp}&=\Lambda(2\ell+1)(\eta_{\rm in}+\eta_{\rm out}).
 \label{eq:fluidmatrix}
\end{align}
For a toroidal amplitude the block is
$a\Lambda[(\ell-1)\eta_{\rm in}+(\ell+2)\eta_{\rm out}]$.
These are the spherical Stokes traction coefficients, with the
tangential-harmonic norm $\Lambda$ included in the work pairing
\citep{Vlahovska2026}. The supplementary material gives a direct
radial-power derivation. The matrix describes an unbounded
linear spherical problem; all computations below concern this
specific bulk-dissipation extension.

\subsection{A modal dominance criterion and rate bounds}
\label{sec:dominance}
Fix a finite harmonic band $2\le\ell\le L_h$ and remove any
zero-strain coordinates. The assembled $K,V$ are SPD and
$R_f\succeq0$. Define
\begin{equation}
 \epsilon_f=\|V^{-1/2}R_fV^{-1/2}\|_2,\qquad
 \delta_f=\frac{\epsilon_f}{1+\epsilon_f}.
 \label{eq:fluidepsilon}
\end{equation}
This dimensionless norm is equivalently the largest generalized
eigenvalue of $(R_f,V)$ and is invariant under a change of basis.
The familiar surface-to-bulk ratios
$\mathrm{Bq}_s=\mu_s/(a\eta_{\rm out})$,
$\mathrm{Bq}_d=\mu_s'/(a\eta_{\rm out})$, together with
$\lambda=\eta_{\rm in}/\eta_{\rm out}$ and the harmonic band,
determine $\epsilon_f$. Both surface coefficients and both fluids
enter this criterion. As an illustrative scale, evaluate the
equal-surface-viscosity inputs of \citet{Guglietta2021},
$\mu_s=\mu_s'=1.59\times10^{-7}$~N\,s\,m$^{-1}$,
$\eta_{\rm in}=6\times10^{-3}$~Pa\,s and
$\eta_{\rm out}=1.2\times10^{-3}$~Pa\,s, on the present
reference sphere at $a=3.5$~$\mu$m. For $\ell=2$ these give
$\mathrm{Bq}_s=\mathrm{Bq}_d\simeq38$ and $\epsilon_f\simeq0.42$,
including both poloidal and toroidal resistance. This illustrative
model scale calls for the coupled branch rather than the controlled
$\epsilon_f\ll1$ limit. The supplied viscosity convention and values come from a
red-blood-cell model; the calculation here uses spherical geometry.

Let $r_j^{(0)}$ and $r_j^{(f)}$ be the increasingly ordered
generalized eigenvalues of $(K,V)$ and $(K,V+R_f)$, counting
multiplicity. The Rayleigh quotient and minimax principle give
\begin{equation}
 \frac{r_j^{(0)}}{1+\epsilon_f}
       \le r_j^{(f)}\le r_j^{(0)} .
 \label{eq:fluidrates}
\end{equation}
Indeed $V\preceq V+R_f\preceq(1+\epsilon_f)V$.
The bound includes repeated reference rates and identifies
$\epsilon_f\ll1$ as the membrane-dominated regime on this fixed band.
It controls ordered rates; material sensitivity also depends on the
modal patterns and their derivatives.

\subsection{Forward, derivative, and projected-inverse control}
\label{sec:coupledcontrol}
Keep $K$, geometry, bulk viscosities, observation, and initial
physical fields fixed while differentiating the two surface
log-viscosities $\theta_k$. Thus $\partial_{\theta_k}R_f=0$.
In fixed elastic-energy coordinates $x=K^{1/2}q$, define
\begin{align}
 H_f&=K^{1/2}(V+R_f)^{-1}K^{1/2},&
 E_f(t)&=e^{-H_ft},\nonumber\\
 H_0&=K^{1/2}V^{-1}K^{1/2},&
 E_0(t)&=e^{-H_0t}.
\end{align}
Set $b=r_{\max}^{(0)}$, $\alpha=r_{\min}^{(0)}/(1+\epsilon_f)$,
and $g_k=\|V^{-1/2}(\partial_{\theta_k}V)V^{-1/2}\|_2$.
For the two log-viscosities in Eq.~(\ref{eq:kv}), $g_k\le1$ because
$0\preceq\partial_{\theta_k}V\preceq V$ for either positive
surface-viscosity contribution.
The following operator bounds hold for $t\ge0$:
\begin{align}
 \|E_f(t)-E_0(t)\|_2
 &\le\delta_f b t e^{-\alpha t},\label{eq:fluidforward}\\
 \|\partial_{\theta_k}E_f(t)-\partial_{\theta_k}E_0(t)\|_2
 &\le\delta_f b g_k(2t+bt^2)e^{-\alpha t}.
 \label{eq:fluidderivative}
\end{align}
They follow from symmetric contraction estimates and Duhamel's
formula, as shown in the supplementary material.
Multiplication by the fixed initial energy-coordinate vector and
the observation $O K^{-1/2}$ converts them into bounds on the
measured response and its material derivatives. Stacking fixed
sample times and whitening with the same SPD weight preserves
these bounds with the corresponding operator norm factors.

For the calibrated pair, let $J_f,J_0$ now denote the whitened
surface-log-viscosity Jacobians and let $n_f,n_0$ denote their
whitened common-amplitude derivatives. The same fixed signed
$\kappa$ prescribes
$q_{\rm tan}(0)=(0,2\kappa C/\Lambda)^{\mathsf T}$ and
$q_{\rm rad}(0)=(C,0)^{\mathsf T}$.
Write $\eta_n=\|n_f-n_0\|_2$ and
$\zeta_J=\|J_f-J_0\|_2$. If $\eta_n<\|n_0\|_2$, then
\begin{align}
 \sigma_{\min}(P_{n_f}^{\perp}J_f)
 &\ge \sigma_{\min}(P_{n_0}^{\perp}J_0)
       -\zeta_J \nonumber\\
 &\quad-\frac{\eta_n}{\|n_0\|_2}\|J_0\|_2 .
 \label{eq:fluidprojected}
\end{align}
The angle between the two nonzero nuisance columns bounds the
projector difference by $\eta_n/\|n_0\|_2$; a singular-value
perturbation inequality completes the result. Conservative
upper bounds on $\eta_n,\zeta_J$ follow from
Eqs.~(\ref{eq:fluidforward}) and (\ref{eq:fluidderivative}).
A positive right-hand side certifies local material rank. For the
coincident-rate design of Fig.~\ref{fig:paired}, with equal inner and
outer fluid viscosities, this sufficient certificate reaches zero near
$\epsilon_f=0.00409$. That value marks the extent of the conservative
bound for this design; beyond it, the full projected Jacobian determines
the sensitivity. Section~\ref{sec:verification} reports that calculation
at finite coupling. On a fixed compact design with amplitude bounded
away from zero, the bound establishes retention through reference-rate
coincidence for sufficiently small $\epsilon_f$.

For $R_f>0$, Eq.~(\ref{eq:pairedtransform}) remains an invertible
data transformation, and inference uses the coupled propagator and
its derivatives to represent the means. The fit uses supplied bulk
viscosities and differentiates the complete response with respect to
surface parameters and any fitted calibration variables. This
preserves the distinction between the reference separation map and
the coupled inverse model.

\subsection{Effect of coupling on calibrated separation}
\label{sec:couplingtradeoff}
\begin{figure*}[t]
 \includegraphics[width=\textwidth]{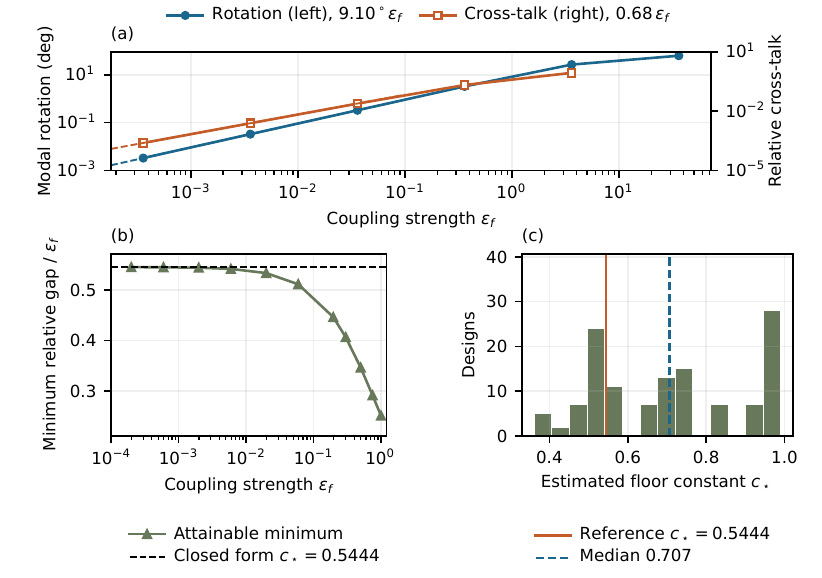}
 \caption{Effect of spherical bulk dissipation on the prescribed-release
 problem. (a) Modal-direction rotation and residual cross-talk after applying
 the membrane-only map (\ref{eq:releasemap}) at separated reference rates.
 Rotation uses the left axis and cross-talk the right axis; vertical
 positions across the two axes are not comparable. Dashed small-coupling
 asymptotes are shown for $\epsilon_f\le0.1$.
 (b) The minimum of $(r_+-r_-)/r_+$ over $\mu_s'$, divided by
 $\epsilon_f$ evaluated at the minimizer, approaching $c_\star$ as coupling decreases.
 The horizontal dashed line is the closed-form $c_\star=0.544444\ldots$.
 (c) Small-coupling numerical estimates of $c_\star$ at
 $\epsilon_f=10^{-4}$ over 126 designs with degrees
 $\ell\in\{2,3,4,6,8,10\}$, fluid viscosity ratios from 0.1 to 10,
 and $c\in\{1,5,11\}$. Panels (a) and (b) use $G_s=1$, $C_a=2$,
 $\mu_s=1.9$, $a=1$, $\ell=2$ and equal fluid viscosities in the
 units of Sec.~\ref{sec:verification}; panel (a) fixes $\mu_s'=1.3$,
whereas panel (b) minimizes over it. In all panels, $\epsilon_f$ is
 evaluated on the one-harmonic poloidal block. The plots describe
 the spectrum and modal map of $(V+R_f)\dot q+Kq=0$; material
 sensitivity is assessed by the full Jacobian.
 Detailed grids and quantitative checks are in the supplementary material.}
 \label{fig:coupled}
\end{figure*}

The membrane forms $K,V\in\operatorname{span}\{R_s,R_d\}$ share a
basis fixed by harmonic degree. This supplies the geometric
release-to-record map (\ref{eq:releasemap}). In the poloidal frame
$T=[\hat q_D\ \hat q_T]$, with $\hat q_D=(1,0)^{\mathsf T}$ and
$\hat q_T=(\Lambda,2)^{\mathsf T}$, membership in that span requires
$\hat q_D^{\mathsf T}X\hat q_T=0$. The Stokes block instead satisfies
\begin{equation}
 \begin{split}
 \hat q_D^{\mathsf T}R_f\hat q_T
 =a\bigl[&(\ell+1)(2\ell+3)(\ell-1)\eta_{\rm in}\\
 &+\ell(\ell+2)(2\ell-1)\eta_{\rm out}\bigr]>0 .
 \end{split}
 \label{eq:transversedefect}
\end{equation}
Thus $R_f$ lies outside the membrane span for every $\ell\ge2$ and
every positive fluid-viscosity pair. Figure~\ref{fig:coupled} shows
the consequences for the modal map and rate separation.

Coupled modal patterns depend on the surface parameters as well as
the supplied fluid coefficients. Applying the membrane-only map
leaves cross-talk. Rotation and cross-talk are $O(\epsilon_f)$ at a
fixed separated-rate reference, but that rotation statement is not
uniform at membrane-only coincidence, where individual modal
directions are degenerate. The full-propagator derivative bound of
Sec.~\ref{sec:coupledcontrol} is the appropriate inverse criterion.

The same transverse term prevents an exact double rate. Generic
one-parameter symmetric pencils avoid repeated eigenvalues
\citep{FriedlandSimon1982}; here the strictly positive expression
(\ref{eq:transversedefect}) proves non-crossing for every admissible
surface-viscosity pair in this one-harmonic model. Write $r_-<r_+$
for the coupled generalized eigenvalues. For an interior minimizer,
the supplementary material derives the minimum at fixed $\mu_s$:
\begin{equation}
 \min_{\mu_s'>0}\frac{r_+-r_-}{r_+}
 =2\nu\left(\sqrt{1+\nu^2}-\nu\right)
 \simeq c_\star\epsilon_f\quad(\epsilon_f\to0),
 \label{eq:gapfloor}
\end{equation}
where $\nu=\mathcal N_{12}/(1+\mathcal N_{22})$, $c=1+2C_a$, and
$\epsilon_f$ is evaluated on this one-harmonic poloidal block.
The modal normalization is
$\Psi=[\Psi_D\ \Psi_T]$,
$\Psi_D=\hat q_D/\sqrt{4c\mu_s}$,
$\Psi_T=\hat q_T/\sqrt{4\mu_s\Lambda(\Lambda-2)}$,
and $\mathcal N=\Psi^{\mathsf T}R_f\Psi$.
Then $\Psi^{\mathsf T}V_0\Psi=I$ for $V_0=\mu_s(R_s+cR_d)$.
The matrix $\mathcal N$ is orthogonally similar to $V_0^{-1/2}R_fV_0^{-1/2}$,
but its individual entries must be taken in the stated modal frame.

The exact minimum in Eq.~(\ref{eq:gapfloor}) requires an interior
minimizer:
$(1+\mathcal N_{22})(1+2\nu^2)>\mathcal N_{11}$.
The reference membrane-dominance condition
$\lambda_{\max}(\mathcal N)\le1$ suffices; interiority can also hold
outside that regime. If interiority fails, the infimum is approached as
$\mu_s'\to0+$. The small-coupling constant is
\begin{equation}
 c_\star=\frac{2\mathcal N_{12}}{\lambda_{\max}(\mathcal N)}
 =\left(1-\frac{\lambda_{\min}(\mathcal N)}
                  {\lambda_{\max}(\mathcal N)}\right)|\sin2\varphi|,
 \label{eq:gapfloorconstant}
\end{equation}
where $\varphi$ is the angle of the resistance eigenbasis relative
to the membrane modal frame. It depends only on $\ell,c$ and the
fluid-viscosity ratio, satisfies $0<c_\star<1$, and equals
$0.544444$ for the reference family.

Non-crossing, modal visibility and material identifiability are
different properties. At $\ell=2$, $c=5$, $G_s=\mu_s=a=1$,
$\eta_{\rm in}=\eta_{\rm out}=\gamma$ and
$\mu_s'=5+15\gamma/8$, a coupled eigenvector is purely tangential
and normal observation annihilates it. Even then, material
derivatives can restore rank through the parameter-dependent patterns.
For $\gamma=1$, unit tangential and radial releases at
$t=(0,0.2,0.4,0.6)$, identity weight and a fitted common amplitude
evaluated at one,
the projected Jacobian with respect to $(\mu_s,\mu_s')$ has singular
values $0.23957$ and $0.0035320$, despite a rank-one observed
modal-amplitude matrix. The supplementary material specifies this
Fr\'echet-derivative check.
Accordingly, the fixed-pattern modal-independence theorem of
Sec.~\ref{sec:independence} belongs to the membrane-only model.
The full Jacobian and Eq.~(\ref{eq:fluidprojected}) assess coupled
material sensitivity, including derivatives of the modal patterns.

\section{Numerical verification}
\label{sec:verification}
The numerical checks compare independently assembled weak-matrix
dynamics with the analytic modes, reconstruct the spherical Stokes
traction, and test the rate and amplitude-projected sensitivities.
The continuum example uses $G_s/G_0=1$, $C_a=2$,
$\mu_s/(G_0t_0)=0.7$ and $\mu_s'/(G_0t_0)=2.1$, with fixed
reference modulus $G_0$ and time $t_0$; plotted time is $t/t_0$.
Modal displacements and scalar signals use the common small displacement
unit $u_0\ll a$, and initial energies use the unit $G_0u_0^2$.
Table~\ref{tab:verification} gives representative discretization errors.
The supplementary material specifies the numerical designs and reports
the detailed verification results.

\begin{table}[t]
\caption{Representative double-precision verification results. Errors
are absolute unless designated relative, and are rounded up to two
significant figures. Quadrature examines
$\ell=2,3,4$; Gauss--Legendre entries use at least eight nodes.
Error entries are numerical residuals of the stated checks, not
uncertainty estimates.}
\label{tab:verification}
\begin{ruledtabular}
\begin{tabular}{lc}
Check & Result\\
\hline
Harmonic forms, Gauss--Legendre relative & $<4\times10^{-15}$\\
Composite Simpson observed order & $3.98$--$4.00$\\
Diagonal modal-ODE amplitudes & $5.2\times10^{-13}$\\
Signed shape-transient rates, relative & $1.7\times10^{-14}$
\\
Independent weak-matrix time solve, relative & $4.6\times10^{-10}$
\end{tabular}
\end{ruledtabular}
\end{table}

The weak-matrix implementation uses fourth-order Runge--Kutta stepping
rather than the modal solution. Hyperbolic series and high-precision
calculations resolve close-rate derivatives without subtracting nearly
equal exponentials; complex-step and Fr\'echet derivatives provide
independent checks. Across the specified schedules and covariances, the
calibrated pair, uncalibrated pair and tangentially excited single record
recover the sensitivity orders in Table~\ref{tab:comparison}.
Numerical positive-gap calculations must preserve the span
of the small amplitude column rather than drop it by a
pseudoinverse tolerance.
At very small gaps a machine-precision forward residual can coexist
with inaccurate recovered parameters. Inverse accuracy is assessed
from the recovered parameters as well as the fit residual.

The coupled survey uses $48$ designs, combining four harmonic degrees,
four fluid-viscosity ratios and three reference rate gaps at one fixed
area parameter, release pair and sampling schedule. Each design has an
uncoupled reference and four nonzero coupling levels, giving $192$
coupled cases. Physical bulk coefficients remain fixed when the surface
viscosities are differentiated. Equation~(\ref{eq:fluidprojected})
certifies all $48$ designs at $\epsilon_f=10^{-3}$ and $20$ at $10^{-2}$.
At $10^{-1}$ and $1$, where that certificate is inconclusive, the
computed full projected Jacobian retains $0.89$--$0.92$ and
$0.43$--$0.54$ of the corresponding membrane-only smallest singular
value, respectively. These finite-grid comparisons complement the
small-coupling analytical bound. A separate deterministic fit illustrates
why the forward model matters: at $\epsilon_f=0.42$, omitting bulk
dissipation gives approximately $42\%$ and $39\%$ errors in the two
viscosities despite a weighted signal residual below $1\%$. Fitting the
coupled model recovers the prescribed values to numerical accuracy
(supplementary material, Sec.~S5).

GPT-6 (OpenAI, accessed through Codex) and Claude Opus 5 (Anthropic)
assisted literature searches, mathematical derivations, code development,
numerical tests and figure-generation code to support the symbolic
analysis and its numerical implementation. The checks above compare
the expressions with separately assembled numerical implementations.
The authors critically reviewed and revised the AI-assisted material.

\section{Discussion}

The release pair separates two sources of ambiguity: nearby decay rates
and unknown excitation amplitudes. In the membrane reference, the radial
record identifies the dilatational contribution, and its calibrated
combination with the tangential-release record isolates the shear
contribution. This separation survives rate coincidence because the
known relative preparation preserves two independent response channels.
The common amplitude is still fitted. Calibration therefore changes
the local sensitivity order, rather than merely improving the signal
level of an otherwise unchanged inverse problem.

The preparation and elastic inputs determine how these rate results
translate into viscosity estimates. The holding tractions specify the
initial material fields; their relative amplitude, gain and timing fix
the linear combination of the records. With $C_a$ supplied, co-fitting
$G_s$ determines the viscosity ratio, whereas absolute viscosities
require an elastic scale. Uncertainty in the relative preparation can be
included by fitting $\kappa$ jointly with the rates and common amplitude
while retaining its independent calibration datum. Under the Gaussian
model in Eq.~(\ref{eq:pairedcalibration}), the full observation covariance
and calibration information then determine the local information bound.
This formulation connects a calibration precision to rate sensitivity
for a specified sampling design.

Bulk dissipation changes both the relaxation spectrum and the modal
patterns, as in the treatments of \citet{Rochal2005} and
\citet{Kuriabova2008,KuriabovaErratum}. The spherical Stokes extension
shows why spectral separation alone is insufficient: material
sensitivities also differentiate the parameter-dependent response
patterns. Its full Jacobian supplies a sufficient small-coupling bound
and a direct numerical test beyond that bound. The deterministic fits
further show that a small signal residual can conceal viscosity bias
when the bulk contribution is omitted. This extension uses known fluid
coefficients and finite area elasticity; a joint estimate of fluid and
surface viscosities, or a local-area-incompressible membrane, defines a
different inverse problem with its own admissible modes and rank test.

\section{Conclusion}
Calibrated, independent release patterns retain two locally distinguishable
viscosity channels even when the intrinsic membrane-relaxation rates
coincide. For fixed sampling and positive-definite weights, bounded
positive rates and nonzero excitation, the paired normal-displacement
records have a positive amplitude-projected sensitivity bound. The
common amplitude remains unknown, and fixed finite initial energy is
compatible with the bound. Independent ratio calibration of fixed
positive precision also retains local sensitivity under the stated
Gaussian model.

The distinction is quantitative: the calibrated pair has order-one
rate sensitivity at coincidence, whereas an uncalibrated pair loses it
linearly and the single tangential-release record loses it quadratically.
For the spherical Stokes extension, the corresponding assessment uses
the full material Jacobian, including changes in modal patterns.
Together, these results identify relative preparation as the information
that removes the coincidence singularity in the membrane reference and
specify how to test its benefit when bulk dissipation is included.

\section*{Supplementary material}
See the supplementary material for the differential-rank and weighting
tools, near-coincidence and gap-floor proofs, strain and Stokes derivations, finite-gap and joint-Jacobian
expansions, and the numerical observation designs and verification
details.

\begin{acknowledgments}
J.O.\ acknowledges support from the University of Illinois Chicago
Chancellor's Undergraduate Research Award (CURA).

GPT-6 (OpenAI, accessed through Codex) and Claude Opus 5 (Anthropic)
assisted manuscript drafting. Their use in the calculations and the use
of these tools in the literature searches are described in
Sec.~\ref{sec:verification}. The authors are responsible for the accuracy
and originality of this work.
\end{acknowledgments}

\section*{Author Declarations}
\subsection*{Conflicts of Interest}
The authors have no conflicts to disclose.

\subsection*{Author Contributions}
\textbf{Julian Olszewski:} Conceptualization (equal), Data curation (equal), Formal analysis (equal),
Funding acquisition (lead), Investigation (equal), Methodology (equal),
Software (equal), Visualization (equal), Writing -- original draft (equal),
Writing -- review \& editing (equal).

\textbf{Zijian Liu:} Conceptualization (equal), Data curation (equal), Formal analysis (equal),
Investigation (equal), Methodology (equal), Software (equal), Visualization (equal),
Writing -- original draft (equal), Writing -- review \& editing (equal).

\textbf{Jie Xu:} Conceptualization (equal), Supervision (lead),
Writing -- review \& editing (equal).

Julian Olszewski and Zijian Liu contributed equally to this work.

\section*{Data Availability}
The data that support the findings of this study are available within the article and its supplementary material.

\bibliography{}
\clearpage
\setcounter{section}{0}
\setcounter{subsection}{0}
\setcounter{equation}{0}
\setcounter{figure}{0}
\setcounter{table}{0}
\renewcommand{\theequation}{S\arabic{equation}}
\renewcommand{\thesection}{S\arabic{section}}
\renewcommand{\thetable}{S\Roman{table}}
\renewcommand{\thefigure}{S\arabic{figure}}
\renewcommand{\theHsection}{S.\arabic{section}}
\renewcommand{\theHsubsection}{\theHsection.\arabic{subsection}}
\renewcommand{\theHequation}{S.\arabic{equation}}
\renewcommand{\theHfigure}{S.\arabic{figure}}
\renewcommand{\theHtable}{S.\arabic{table}}
\pdfbookmark[0]{Supplementary Material}{supplementary-material}
\begin{center}
{\Large\bfseries Supplementary Material\par}
\vspace{0.6em}
{\large Calibrated shape-only recovery of membrane viscosities\par}
\end{center}

Section and equation numbers carrying an S prefix belong to this
document. All other numbers refer to the article.

\section{Differential rank and the effect of weighting}\label{si-app:rank}
This section expands the sensitivity definition in
Sec.~\ref{sec:sensitivity} of the article. The references on
local identifiability, identifiable combinations, and variable
projection are cited there and listed in the article's bibliography.

For a differentiable response $z(\vartheta,\nu)$, let
$J=\partial z/\partial\log\vartheta$ and
$N=\partial z/\partial\nu$ be the material and total nuisance
derivatives, with $p$ material and $q$ nuisance coordinates.
The Euclidean projected matrix satisfies
\begin{equation}
 M=P_N^\perp J,\qquad
 \rank M=\rank[N,J]-\rank N ,
 \label{si-eq:increment}
\end{equation}
where $P_N^\perp=I-NN^\dagger$.
For a composition $z=\mathcal Z(\beta(\vartheta,\nu),\nu)$,
the chain rule also gives
\begin{equation}
 M=\Xi\,\Gamma\,D,\quad
 \Xi=P_N^\perp\partial_\beta\mathcal Z,\quad
 \Gamma=\partial_\vartheta\beta,\quad D=\diag\vartheta .
 \label{si-eq:main}
\end{equation}
The total $N$ includes nuisance dependence through $\beta$.
Full material rank means that no first-order material change
can be canceled by a nuisance change.

If $N$ has full column rank $q$ and $\rank M=p$, the joint
Jacobian has rank $p+q$. A nonsingular square output minor
and the inverse function theorem give a regular local joint
inverse. Singular nuisance coordinates require separate exact
analysis. For example, $z=\vartheta-(\nu-1)^2$ at $(1,1)$ has
$N=0,J=1$ but the curve $\vartheta=1+(\nu-1)^2$ preserves
the observation. Conversely, $z=(\vartheta-1)^3$ is locally
injective despite a zero derivative at $\vartheta=1$. Thus differential
rank and exact identifiability are distinct statements.

The weighted projected matrix and its residual characterization are
given by Eqs.~(\ref{eq:weighted}) and
(\ref{eq:residual}) of the article. Invertible whitening
preserves incremental rank.
Bounding the quadratic form before minimization by the extreme
eigenvalues of $W$ gives
\begin{equation}
 \lambda_{\min}(W)\,M^{\mathsf T}M\preceq M_W^{\mathsf T}M_W
                  \preceq\lambda_{\max}(W)\,M^{\mathsf T}M .
 \label{si-eq:bound}
\end{equation}
For full column rank $M$,
\begin{equation}
 \max\left\{1,\frac{\operatorname{cond}(M)}{\sqrt{\operatorname{cond}(W)}}\right\}
 \le\operatorname{cond}(M_W)\le\sqrt{\operatorname{cond}(W)}\,\operatorname{cond}(M).
 \label{si-eq:kappa}
\end{equation}
This comparison holds for fixed observation design, nuisance
set, and material coordinates. The eigenvalues of $W$ bound
the change, but its eigenvectors relative to the sensitivities
determine the actual conditioning. Therefore $\operatorname{cond}(W)$
alone does not specify the projected condition number.
The numerical weights are prescribed; a likelihood
interpretation is used only for the explicitly stipulated
calibration model in Sec.~\ref{sec:paired}.

\subsection{Sharpness of the coincidence bracket}
\label{si-app:coincidencesharpness}
Fix a common schedule of $n\ge2$ distinct times, $r>0$, $S\ne0$, and
the positive-definite factors $W_r,W_t$ of
Eq.~(\ref{eq:coincidencebracket}). Let
$e=(e^{-rt_j})_j$, $v=(-t_je^{-rt_j})_j$ and
$s_\theta=\sin\angle_{W_t}(v,e)$. Two or more distinct times make $e,v$
independent, so $s_\theta>0$, whereas a single time gives $s_\theta=0$;
$s_\theta$ depends on $r$ through $e$ and $v$. Set $H=W_r^{1/2}\mathcal M$,
$h=(1,1)^{\mathsf T}/\sqrt2$, $k=(1,-1)^{\mathsf T}/\sqrt2$ and
$U=[h\ k]$. The whitened amplitude column is proportional to
$(Hh)\otimes(W_t^{1/2}e)$. Projecting it out gives
\begin{equation*}
 \begin{aligned}
 \frac{M_\kappa^{\mathsf T}M_\kappa}{S^2\|v\|_{W_t}^2}
 &=H^{\mathsf T}\left[I-(1-s_\theta^2)\hat n\hat n^{\mathsf T}\right]H,\\
 \hat n&=Hh/\|Hh\|.
 \end{aligned}
\end{equation*}
For $H=\diag(1,b)U^{\mathsf T}$ with $b\ge1$, the right-hand side
has eigenvalues $s_\theta^2,b^2$. Since $\sigma_{\min}(H)=1$, this
attains the lower bracket endpoint. For
$H=\diag(b,1)U^{\mathsf T}$ with $b\ge1/s_\theta$, its eigenvalues
are $b^2s_\theta^2,1$, attaining the upper endpoint. In both cases
$\mathcal M=W_r^{-1/2}H$ is real and nonsingular. The invertible
release map (\ref{eq:releasemap}) realizes each matrix with
prescribed poloidal releases. A common rescaling keeps the
displacements small without changing either ratio. Thus both factors
are optimal for every fixed common schedule of at least two distinct
times and every factorized weight.

\section{Pattern dependence of the amplitude-projected coincidence order}
\label{si-app:pattern}
Write a single normal record in the form of
Eq.~(\ref{eq:generalrecords}),
\begin{equation}
 y(t)=S\left(Pe^{-r_Tt}+Qe^{-r_Dt}\right),
 \label{si-eq:patrecord}
\end{equation}
with $(P,Q)$ known from the prescribed release through
Eq.~(\ref{eq:releasemap}) and $S\ne0$ the one profiled amplitude.
Throughout this section, use the local full-gap convention
$r_T=r+d/2$, $r_D=r-d/2$, so that $d=r_T-r_D>0$ on the branch treated
here. Write $h=d/2$ for the article's half-gap in
Sec.~\ref{sec:collision}; the conversion of sensitivities and
condition numbers is given after Eq.~(\ref{si-eq:patB}).
The branch $r_T<r_D$ is the same record
with $P$ and $Q$ exchanged, so the results below apply to it with
$(P,Q)$ replaced by $(Q,P)$. Material coordinates are $(r,d)$
and the nuisance coordinate is $S$. Set
\begin{equation}
 s=P+Q,\qquad \delta=Q-P,\qquad s^2-\delta^2=4PQ,
 \label{si-eq:patsd}
\end{equation}
and note the identification $y(0)=Ss$: the pattern sum is the initial
normal signal. Throughout $n_S=\partial y/\partial S$,
$J=[\partial y/\partial r,\ \partial y/\partial d]$ and
$M_W=P^{\perp}_{Ln_S}LJ$ in the notation of
Sec.~\ref{si-app:rank}.

Set $e=(e^{-rt_j})_j$ and $te=(t_je^{-rt_j})_j$.
With the full-gap convention of this section, the fitted-amplitude
derivative has the expansion
\begin{equation}
 n_S=(P+Q)\,e-\tfrac{1}{2}(P-Q)\,d\,te+O(d^2).
 \label{si-eq:nuisanceexpansion}
\end{equation}
Thus whether $P+Q$ vanishes selects the leading direction $te$ or $e$.
The lemmas below quantify how fitting that amplitude changes
the weakest rate sensitivity.

The hypotheses are: (H1) $S\ne0$ and $r>0$ fixed, $0<d<d_0$;
(H2) $W=L^{\mathsf T}L$ fixed and positive definite with $L$
invertible, independent of $d$; (H3) fixed nonnegative sample times
$t_1<\dots<t_n$, independent of $d$, at least three of them distinct,
and for the case $s=0$ at least three of them distinct and positive;
(H4) $PQ\ne0$. When $s=0$, this is equivalent to $P\ne0$ because $Q=-P$.

Hypothesis (H4) is not cosmetic. If $Q=0$ then
$y=SPe^{-(r+d/2)t}$ depends on $(r,d)$ only through the single
combination $r+d/2$, so $\rank J\le1$ and
$\sigma_{\min}(M_W)=0$ identically, for every $d$ and every schedule;
likewise for $P=0$. Both channels must be excited before any gap
order can be discussed. Define the weighted moment vectors
$\chi_k=(t_j^ke^{-rt_j})_j$ and $x_k=L\chi_k$, which are the vectors
of Sec.~\ref{sec:collision} with $m=r$, and the residuals
\begin{equation}
 \begin{aligned}
  a_1&=P^{\perp}_{x_0}x_1, &\quad b_2&=P^{\perp}_{[x_0,x_1]}x_2,\\
  a_2&=P^{\perp}_{x_1}x_2, &\quad b_3&=P^{\perp}_{[x_1,x_2]}x_3 .
 \end{aligned}
 \label{si-eq:patresiduals}
\end{equation}

\subsection{Supporting lemmas}
\emph{Lemma 1.} With $t\odot w$ the elementwise product and
\begin{equation}
 \begin{aligned}
  u(d)&=\left(e^{-rt_j}\left[s\cosh\tfrac{dt_j}{2}
        +\delta\sinh\tfrac{dt_j}{2}\right]\right)_j,\\
  v(d)&=\left(e^{-rt_j}\left[\delta\cosh\tfrac{dt_j}{2}
        +s\sinh\tfrac{dt_j}{2}\right]\right)_j,
 \end{aligned}
 \label{si-eq:patuv}
\end{equation}
one has exactly $n_S=u$, $\partial y/\partial r=-S\,t\odot u$ and
$\partial y/\partial d=\tfrac{1}{2}S\,t\odot v$, with the entire
expansions
\begin{equation}
 \begin{aligned}
  n_S&=\sum_{k\ge0}c^u_k\chi_k, &\quad
  \frac{\partial y}{\partial r}&=-S\sum_{k\ge0}c^u_k\chi_{k+1},\\
  \frac{\partial y}{\partial d}&=\frac{S}{2}\sum_{k\ge0}c^v_k\chi_{k+1},
 \end{aligned}
 \label{si-eq:patseries}
\end{equation}
where $c^u_k=(d/2)^k[Q+(-1)^kP]/k!$ and
$c^v_k=(d/2)^k[Q-(-1)^kP]/k!$. The identity
$s\cosh\xi+\delta\sinh\xi=Qe^{\xi}+Pe^{-\xi}$ with $\xi=dt/2$ gives
Eq.~(\ref{si-eq:patuv}); differentiation gives the three columns; the
series follow from $t^k\odot\chi_0=\chi_k$ and
$t\odot\chi_k=\chi_{k+1}$, and converge for every $d$ because the
exponentials are entire. Three structural facts are worth retaining:
the two rate columns are $t$ times $u$ and $v$; the nuisance column
\emph{is} $u$; and in the moment frame the rate columns are the
nuisance-type expansions shifted one rung up the ladder
$\chi_k\mapsto\chi_{k+1}$.

\emph{Lemma 2.} For an $n\times2$ matrix $M$ write
$V_2(M)=(\det M^{\mathsf T}M)^{1/2}$, and write $V_3$ for the same
quantity formed from three columns, again as a square root. Then
\begin{equation}
 \begin{aligned}
  \sigma_{\min}(M)\,\sigma_{\max}(M)&=V_2(M),\\
  V_3(c_0,c_1,c_2)&=\|c_0\|\,
      V_2(P^{\perp}_{c_0}c_1,P^{\perp}_{c_0}c_2),
 \end{aligned}
 \label{si-eq:gramfactor}
\end{equation}
the second for $c_0\ne0$. Combining the two relations of
Eq.~(\ref{si-eq:gramfactor}),
\begin{equation}
 \sigma_{\min}(M_W)=
 \frac{V_3\!\left(Ln_S,L\partial_ry,L\partial_dy\right)}
      {\|Ln_S\|\;\sigma_{\max}(M_W)} .
 \label{si-eq:patkey}
\end{equation}
The squared numerator of Eq.~(\ref{si-eq:patkey}) is a real-analytic
Gram determinant in $d$, with no projector. The leading
exterior-product term for $d>0$ gives the volume's asymptotic
with the nuisance direction varying with $d$. Only $\sigma_{\max}$
still involves the projector, and a leading-order direction suffices because
$\sigma_{\max}$ is needed only to relative accuracy $O(d)$.

\emph{Lemma 3.} Under (H2), $\{x_0,x_1,x_2\}$ is independent if and
only if the schedule has at least three distinct times, and
$\{x_1,x_2,x_3\}$ if and only if it has at least three distinct
\emph{positive} times. In the first case,
$V_3(x_0,x_1,x_2)=\|x_0\|\|a_1\|\|b_2\|$ with $\|a_1\|,\|b_2\|>0$.
In the second,
$V_3(x_1,x_2,x_3)=\|x_1\|\|a_2\|\|b_3\|$ with $\|a_2\|,\|b_3\|>0$. Writing $\chi_k=\Delta(t_j^k)_j$ with
$\Delta=\diag(e^{-rt_j})$ invertible, independence reduces to that of
the columns of the Vandermonde matrix with rows $(1,t_j,t_j^2)$,
respectively $t_j(1,t_j,t_j^2)$, in which a row with $t_j=0$ is
identically zero. Two consequences follow at once: at most $n-1$
directions survive the projection of one nuisance column, so $n=2$
gives $\sigma_{\min}(M_W)=0$ for every pattern; and in the shifted
ladder a sample at $t=0$ contributes nothing at any order.

\emph{Lemma 4.} Let $m_{ijk}(d)$ be the coefficients of
$(Lu)\wedge(L\,t\odot u)\wedge(L\,t\odot v)$ in the moment frame.
Then exactly
\begin{equation}
 m_{012}=2PQ(P+Q)\,d,\qquad m_{013}=0,
 \label{si-eq:patminors}
\end{equation}
and every other $m_{ijk}$ is $O(d^3)$. If $P+Q=0$ then
$m_{123}=-P^3d^4/6$ exactly and every other $m_{ijk}$ vanishes or is
$O(d^6)$. By Lemma 1 the three coefficient rows are $(c^u_k)_{k\ge0}$,
$(0,c^u_0,c^u_1,\dots)$ and $(0,c^v_0,c^v_1,\dots)$. For the triple
$(0,1,k)$ the first column has the single entry $c^u_0=s$, so
\begin{equation}
 m_{01k}=s\,\frac{(d/2)^{k-1}}{(k-1)!}\,2PQ\left[1+(-1)^k\right],
 \label{si-eq:patm01k}
\end{equation}
which vanishes for odd $k$ and gives Eq.~(\ref{si-eq:patminors}) at
$k=2$. The remaining minors are direct determinants of those rows, in
which every entry beyond the first column of the shifted rows carries
an explicit power of $d$. It is Eq.~(\ref{si-eq:patsd}) that produces the
factor $PQ$, and the factor $s$ in Eq.~(\ref{si-eq:patm01k}) that
collapses the whole family $(0,1,k)$ when the initial normal signal
vanishes.

\subsection{Generic pattern: linear coincidence order}
\emph{Proposition 1.} Assume (H1) to (H4) with $s\ne0$. Then
$Ln_S=sx_0+O(d)$ is bounded away from zero, the unit vector
$Ln_S/\|Ln_S\|$ is real analytic at $d=0$ with limit
$\operatorname{sign}(s)\,x_0/\|x_0\|$,
and
\begin{equation}
 \begin{aligned}
  P^{\perp}_{Ln_S}L\partial_ry&=-Ss\,a_1+O(d),\\
  P^{\perp}_{Ln_S}L\partial_dy&=\tfrac{1}{2}S\delta\,a_1+O(d),
 \end{aligned}
 \label{si-eq:patAcols}
\end{equation}
so the limiting projected block is rank one along $a_1$, with
\begin{equation}
 \sigma_{\max}(M_W)=\tfrac{1}{2}|S|\,\|a_1\|
 \sqrt{4s^2+\delta^2}+O(d)
 \label{si-eq:patAmax}
\end{equation}
and
\begin{equation}
 \sigma_{\min}(M_W)=
 \frac{2|S|\,|PQ|\,\|b_2\|}{\sqrt{4s^2+\delta^2}}\;d+O(d^2)
 =\Theta(d).
 \label{si-eq:patAmin}
\end{equation}
The nearly unidentifiable rate direction tends to the direction
parallel to $\left(\tfrac{1}{2}\delta,\,s\right)$ in $(r,d)$.
Equation~(\ref{si-eq:patAcols}) is Lemma 1 truncated at $k=0$, and
Eq.~(\ref{si-eq:patAmax}) is the largest singular value of
$[-Ss\,a_1,\ \tfrac12S\delta\,a_1]$. For the smallest value, the
three columns of Eq.~(\ref{si-eq:patkey}) give
$V_3=S^2|PQ\,s|\,d\,\|x_0\|\|a_1\|\|b_2\|+O(d^3)$ by Lemma 4 and
Lemma 3, while $\|Ln_S\|=|s|\|x_0\|+O(d)$; dividing by both leaves
Eq.~(\ref{si-eq:patAmin}), with $\|a_1\|$ canceling. Three distinct
times suffice, and one of them may be $t=0$: that sample returns
$y(0)=Ss$, so a schedule $\{0,t_2,t_3\}$ already gives
$\sigma_{\min}(M_W)>0$ with the same constant.

\subsection{Equal-and-opposite pattern: quadratic coincidence order}
Now let $s=0$ with $P\ne0$, equivalently $y(0)=0$.
Equation~(\ref{si-eq:patuv}) collapses to
$u=-2P(e^{-rt_j}\sinh\tfrac{dt_j}{2})_j$ and
$v=-2P(e^{-rt_j}\cosh\tfrac{dt_j}{2})_j$, which is
Eq.~(\ref{eq:gapresponse}) with ordered amplitude $B=-SP$, mean rate
$m=r$ and half-gap $d/2$. The three whitened columns are
\begin{equation}
 \begin{aligned}
  Ln_S&=-P\left(d\,x_1+\tfrac{1}{24}d^3x_3+O(d^5)\right),\\
  L\partial_ry&=SP\left(d\,x_2+\tfrac{1}{24}d^3x_4+O(d^5)\right),\\
  L\partial_dy&=-SP\left(x_1+\tfrac{1}{8}d^2x_3+O(d^4)\right).
 \end{aligned}
 \label{si-eq:patBcols}
\end{equation}
Here $Ln_S$ vanishes at $d=0$, so the map $N\mapsto I-NN^{\dagger}$ is
genuinely discontinuous there, as noted in
Sec.~\ref{sec:collision}: the nuisance rank drops from one to zero
and the projector jumps to the identity. The discontinuity is
nevertheless removable on $d>0$. Factoring the scalar $-Pd$, which
changes only the sign of the unit vector, leaves
\begin{equation}
 \hat w(d)=\frac{x_1+\tfrac{1}{24}d^2x_3+O(d^4)}
                {\left\|x_1+\tfrac{1}{24}d^2x_3+O(d^4)\right\|},
 \label{si-eq:patBhatw}
\end{equation}
which is real analytic and even in $d$ with $\hat w(0)=x_1/\|x_1\|$,
so $P^{\perp}_{Ln_S}=I-\hat w\hat w^{\mathsf T}$ extends analytically
through the origin with limit $P^{\perp}_{x_1}$ and
$\|P^{\perp}_{Ln_S}-P^{\perp}_{x_1}\|_2=O(d^2)$.

The projector $P^\perp_{Ln_S}=I-\hat w\hat w^{\mathsf T}$
annihilates $Ln_S$, hence $x_1+\tfrac{1}{24}d^2x_3$ up to $O(d^4)$.
The gap column carries $x_1+\tfrac18d^2x_3$, leaving the coefficient
$\tfrac18-\tfrac1{24}=\tfrac1{12}$:
\begin{equation}
 P^{\perp}_{Ln_S}L\partial_dy
 =-\frac{SP}{12}d^2\,P^{\perp}_{x_1}x_3+O(d^4),
 \label{si-eq:patBdcol}
\end{equation}
while $P^{\perp}_{Ln_S}L\partial_ry=SP\,d\,a_2+O(d^3)$ is insensitive
to it. Freezing the projector at its limit would retain $\tfrac18$ and
overstate the constant by exactly $3/2$, with the correct exponent.

\emph{Proposition 2.} Assume (H1) to (H4) with $s=0$.  Then
\begin{equation}
 \begin{aligned}
  \sigma_{\max}(M_W)&=|SP|\,\|a_2\|\,d+O(d^3),\\
  \sigma_{\min}(M_W)&=\tfrac{1}{12}|SP|\,\|b_3\|\,d^2+O(d^4),
 \end{aligned}
 \label{si-eq:patB}
\end{equation}
and the weak direction tends to the pure gap direction. The two
projected columns have sizes $\Theta(d)$ and $\Theta(d^2)$, which
gives $\sigma_{\max}$; for $\sigma_{\min}$, Lemma 4 and Lemma 3 give
$V_3=\tfrac{1}{12}S^2|P|^3d^4\|x_1\|\|a_2\|\|b_3\|+O(d^6)$ while
$\|Ln_S\|=|P|d\|x_1\|(1+O(d^2))$, and again $\|a_2\|$ cancels.
Writing $h=d/2$ and using rate coordinates $(r,h)$ multiplies the
gap column by two and gives
$\sigma_{\max}=2|SP|\|a_2\|h+O(h^3)$ and
$\sigma_{\min}=(2/3)|SP|\|b_3\|h^2+O(h^4)$,
reproducing Eq.~(\ref{eq:collisionlaw}) with $B=-SP$ and the
article's half-gap $d=h$. In these half-gap coordinates the condition
number is asymptotic to $3\|a_2\|/(\|b_3\|h)$, compared with
$12\|a_2\|/(\|b_3\|d)$ in the full-gap coordinates below.
The coordinate rescaling changes the constants but preserves the powers.
The schedule must contain three distinct \emph{positive}
times: all three columns of Eq.~(\ref{si-eq:patBcols}) carry the factor
$t$, so a row with $t_j=0$ is identically zero and a schedule
$\{0,t_2,t_3\}$ leaves $\sigma_{\min}(M_W)=0$.

\subsection{What the constants depend on}
In both propositions the leading constant factors into an amplitude
factor and one design scalar, and that scalar is a Schur complement of
the weighted Gram matrix of the moment vectors: with
$G_{kl}=\chi_k^{\mathsf T}W\chi_l$,
\begin{equation}
 \begin{aligned}
  \|b_2\|^2&=\frac{\det G(\chi_0,\chi_1,\chi_2)}
                   {\det G(\chi_0,\chi_1)},\\
  \|b_3\|^2&=\frac{\det G(\chi_1,\chi_2,\chi_3)}
                   {\det G(\chi_1,\chi_2)}.
 \end{aligned}
 \label{si-eq:patschur}
\end{equation}
The weight, window and mean rate enter the leading
$\sigma_{\min}$ only through $\|b_2\|$ or $\|b_3\|$;
$\|a_1\|$ and $\|a_2\|$ cancel. The residual
characterization Eq.~(\ref{eq:residual}) gives the sandwich
\begin{equation}
 \begin{split}
  \sqrt{\lambda_{\min}(W)}\,\sigma_{\min}(M_I)
  &\le\sigma_{\min}(M_W)\\
  &\le\sqrt{\lambda_{\max}(W)}\,\sigma_{\min}(M_I),
 \end{split}
 \label{si-eq:patsandwich}
\end{equation}
so the leading constant is the unweighted one times a factor between
$\sqrt{\lambda_{\min}(W)}$ and $\sqrt{\lambda_{\max}(W)}$, and the exponent
never moves. It is the spread of that factor, not its value, that
$\operatorname{cond}(W)^{1/2}$ bounds: $W=100I$ has condition number one and
still multiplies $\sigma_{\min}$ by ten. The
constants degenerate as the schedule approaches fewer than three
admissible distinct times. As in Sec.~\ref{sec:collision}, none of
this is uniform over windows or weights that themselves vary with $d$.

The two cases share their condition-number order,
\begin{equation}
 \operatorname{cond}(M_W)\sim
 \begin{cases}
  \dfrac{\|a_1\|\left(4s^2+\delta^2\right)}{4|PQ|\,\|b_2\|\,d},
   & s\ne0,\\[2ex]
  \dfrac{12\,\|a_2\|}{\|b_3\|\,d}, & s=0,
 \end{cases}
 \label{si-eq:patkappa}
\end{equation}
both $\Theta(d^{-1})$. The condition number therefore does not see the
pattern dichotomy at all; only the absolute projected sensitivity
does. The joint orders are $\Theta(d)$ for $s\ne0$ and
$|P|\|b_3\|d^3/12+O(d^5)$ for $s=0$, consistent with
Table~\ref{tab:comparison} and Sec.~\ref{si-app:comparison}.

These are fixed-pattern asymptotics. For a small nonzero pattern sum,
Proposition 1 applies on an interval that can shrink as $s\to0$;
its finite leading coefficient alone does not specify the joint
small-sum, small-gap limit.

Lemma 3 counts samples for differential rank. Global recovery is
pattern dependent: the finite searches specified in
Sec.~\ref{si-app:designs} exhibit two admissible solutions for three
generic patterns and one for the equal-and-opposite control.
The unique generic example there is established by exact root isolation.
For the equal-and-opposite pattern, the analytic argument of
Sec.~\ref{sec:collision} guarantees recovery from three positive
samples. Its zero initial signal supplies that extra constraint while
reducing the weak projected sensitivity from $\Theta(d)$ to
$\Theta(d^2)$; both condition numbers remain $\Theta(d^{-1})$.

\section{Analytic gap floor and the proportionality constant}
\label{si-app:gapfloor}
Fix $\ell\ge2$, set $\Lambda=\ell(\ell+1)$,
$s=\Lambda(\Lambda-2)>0$ and $c=1+2C_a>0$, and keep
$K=G_s(R_s+cR_d)$, $V=\mu_sR_s+\mu_s'R_d$ with
$G_s,\mu_s,\mu_s'>0$. Write $r_-\le r_+$ for the generalized
eigenvalues of the coupled pencil $(K,D)$ with $D=V+R_f$, so that the
relative gap of Eq.~(\ref{eq:gapfloor}) is $(r_+-r_-)/r_+$. The
Stokes block is symmetric positive semidefinite, and positive definite
whenever at least one bulk viscosity is nonzero.

\subsection{The membrane modal frame diagonalizes the whole span}
Both membrane generators of Eq.~(\ref{eq:poloidalmatrices}) are rank
one, $R_d=vv^{\mathsf T}$ with $v=(2,-\Lambda)^{\mathsf T}$ and
$R_s=s\,ww^{\mathsf T}$ with $w=(0,1)^{\mathsf T}$ and $s=\Lambda(\Lambda-2)$,
which is local to this section. Let $T$ be the
matrix whose columns are the two parameter-independent modal
directions of Sec.~\ref{sec:independence},
\begin{equation}
 T=\begin{bmatrix}\hat q_D&\hat q_T\end{bmatrix}
  =\begin{pmatrix}1&\Lambda\\0&2\end{pmatrix},
 \qquad \det T=2 .
 \label{si-eq:gfcongruence}
\end{equation}
Because $v^{\mathsf T}\hat q_T=0$ and $w^{\mathsf T}\hat q_D=0$, the
congruence $X\mapsto\widetilde X=T^{\mathsf T}XT$ sends both
generators to diagonal form,
$\widetilde R_d=\diag(4,0)$ and $\widetilde R_s=\diag(0,4s)$, hence
\begin{equation}
 \widetilde K=4G_s\diag(c,s),\qquad
 \widetilde V=4\diag(\mu_s',\mu_ss).
 \label{si-eq:gfKV}
\end{equation}
Equation~(\ref{si-eq:gfKV}) is the structural fact of
Sec.~\ref{sec:couplingtradeoff} in its sharpest form: in the frame
$T$ the whole of $\operatorname{span}\{R_s,R_d\}$ is the space of
\emph{diagonal} symmetric forms. Since the off-diagonal entry of
$\widetilde X$ is $\hat q_D^{\mathsf T}X\hat q_T$ and
$\dim\mathrm{Sym}(2)=3$, this gives the coordinate-free
characterization
\begin{equation}
 \operatorname{span}\{R_s,R_d\}
 =\left\{X=X^{\mathsf T}:\ \hat q_D^{\mathsf T}X\hat q_T=0\right\}.
 \label{si-eq:gfspan}
\end{equation}
Generalized eigenvalues are invariant under the congruence, so we may
work with $\widetilde K,\widetilde D$ throughout.

\subsection{The transverse component of the Stokes block never vanishes}
Write $\widetilde R_f=T^{\mathsf T}R_fT$ with entries
$f_{11}=a\rho_{rr}$,
$f_{12}=a(\Lambda\rho_{rr}+2\rho_{rp})$ and
$f_{22}=a(\Lambda^2\rho_{rr}+4\Lambda\rho_{rp}+4\rho_{pp})$. By
Eq.~(\ref{si-eq:gfspan}), $R_f$ lies in the span if and only if
$f_{12}=0$. Substituting Eq.~(\ref{eq:fluidmatrix}) and collecting the
two viscosities,
\begin{align}
 f_{12}&=a\Bigl\{\left[\Lambda(2\ell+1+3/\ell)-6(\ell+1)\right]
  \eta_{\rm in}\nonumber\\
 &\qquad+\left[\Lambda(2\ell+1+3/(\ell+1))-6\ell\right]
  \eta_{\rm out}\Bigr\}\nonumber\\
 &=a\left[(\ell+1)(2\ell+3)(\ell-1)\eta_{\rm in}
  +\ell(\ell+2)(2\ell-1)\eta_{\rm out}\right],
 \label{si-eq:gfdefect}
\end{align}
which is Eq.~(\ref{eq:transversedefect}). Both bracketed integers are
strictly positive for every $\ell\ge2$, the smallest pair being
$(21,24)$ at $\ell=2$. Hence $f_{12}>0$ identically.

\subsection{Discriminant splitting}
The characteristic polynomial of the pencil is
$\det(\widetilde K-r\widetilde D)
=r^2\det\widetilde D-rb+\det\widetilde K$ with
$b=\tr\widetilde K\tr\widetilde D-\tr(\widetilde K\widetilde D)>0$.
Abbreviating $k_1=4G_sc$, $k_2=4G_ss$, $A=4\mu_s'+f_{11}$,
$C=4\mu_ss+f_{22}$ and $F=f_{12}$, the discriminant
$\Delta=b^2-4\det\widetilde K\det\widetilde D$ collapses to
\begin{equation}
 \Delta=(k_1C-k_2A)^2+4k_1k_2F^2 .
 \label{si-eq:gfsplit}
\end{equation}
The two terms are the in-span and the transverse contributions. The
first depends on both surface viscosities and can be annihilated. The
second is \emph{independent of both}, because
$F=\hat q_D^{\mathsf T}R_f\hat q_T$ involves only the Stokes block, and by
Eq.~(\ref{si-eq:gfdefect}) it is strictly positive. Therefore
\begin{equation}
 \Delta\ \ge\ 4k_1k_2F^2>0
 \label{si-eq:gfdeltafloor}
\end{equation}
for every admissible $(\mu_s,\mu_s')$, so $r_-<r_+$ always and exact
coincidence is algebraically unreachable at nonzero coupling. This
part needs no smallness hypothesis and no minimization. In the
membrane-only case $F=0$ and $\Delta=(k_1C-k_2A)^2$ vanishes
identically at $\mu_s'=c\mu_s$, where $V=(\mu_s/G_s)K$, recovering
$r_T=r_D$.

\subsection{Exact minimization over the dilatational viscosity}
From $r_\pm=(b\pm\sqrt\Delta)/(2\det\widetilde D)$ and $b>0$ the
relative gap is $2/(1+b/\sqrt\Delta)$, so minimizing it is maximizing
$b/\sqrt\Delta$. Put $x=k_2A$, $p=k_1C$ and $m^2=4k_1k_2F^2$, so that
$b/\sqrt\Delta=(p+x)/\sqrt{(p-x)^2+m^2}$ by
Eq.~(\ref{si-eq:gfsplit}). Only $x$ depends on $\mu_s'$, affinely and
increasingly, and the stationarity condition reduces to
$2p^2-2px+m^2=0$, that is to the unique critical point
$x_\star=p+m^2/(2p)$, which is a maximum. Setting
\begin{equation}
 \nu=\frac{m}{2p}=\frac{f_{12}}{4\mu_ss+f_{22}}\sqrt{\frac{s}{c}}
 \label{si-eq:gfnu}
\end{equation}
gives $b/\sqrt\Delta=\nu^{-1}\sqrt{1+\nu^2}$ and hence the closed form
of Eq.~(\ref{eq:gapfloor}),
\begin{equation}
 \min_{\mu_s'>0}\frac{r_+-r_-}{r_+}
 =\frac{2\nu}{\nu+\sqrt{1+\nu^2}}
 =2\nu\left(\sqrt{1+\nu^2}-\nu\right),
 \label{si-eq:gfexact}
\end{equation}
attained at
\begin{equation}
 \mu_s'^\star=c\mu_s\left[(1+\mathcal N_{22})(1+2\nu^2)-\mathcal N_{11}\right],
 \label{si-eq:gfminimizer}
\end{equation}
where the three dimensionless resistance coefficients are
\begin{equation}
 \mathcal N_{11}=\frac{f_{11}}{4c\mu_s},\qquad
 \mathcal N_{12}=\frac{f_{12}}{4\mu_s\sqrt{cs}},\qquad
 \mathcal N_{22}=\frac{f_{22}}{4\mu_ss},
 \label{si-eq:gfabg}
\end{equation}
and $\nu=\mathcal N_{12}/(1+\mathcal N_{22})$. Equation~(\ref{si-eq:gfexact}) holds at finite
coupling strength when the maximizer is interior, $\mu_s'^\star>0$,
that is
\begin{equation}
 (1+\mathcal N_{22})(1+2\nu^2)>\mathcal N_{11} ,
 \label{si-eq:gfinterior}
\end{equation}
which holds whenever $\lambda_{\max}(\mathcal N)\le1$ below, since
$\mathcal N_{11}=e_1^{\mathsf T}\mathcal N e_1\le\lambda_{\max}(\mathcal N)$ and $\mathcal N_{22}>0$. The
condition is not universal. Over a deliberately extreme grid of
$1500$ designs, with degrees to $\ell=14$, area parameters down to
$c=0.2$, $\mu_s$ over six decades and both bulk viscosities over
twelve, it fails in nine cases, all of them with
$\lambda_{\max}(\mathcal N)\ge8.8$, and in none with $\lambda_{\max}(\mathcal N)\le1$.
If interiority fails, the unconstrained optimum has $\mu_s'\le0$, and the
infimum is approached as $\mu_s'\to0+$. The membrane
dominance already assumed in Sec.~\ref{sec:dominance} is therefore
sufficient.

\subsection{The proportionality constant}
The coefficients of Eq.~(\ref{si-eq:gfabg}) are exactly the entries of
the Stokes block normalized by the membrane dissipation \emph{at the
coincidence point}. At $\mu_s'=c\mu_s$ one has
$V_0=\mu_s(R_s+cR_d)=(\mu_s/G_s)K$, and the $V_0$-orthonormal modal
frame is $\Psi=[\,\Psi_D\ \ \Psi_T\,]$ with
$\Psi_D=\hat q_D/\sqrt{4c\mu_s}$ and $\Psi_T=\hat q_T/\sqrt{4\mu_ss}$,
so that $\Psi^{\mathsf T}V_0\Psi=I$ and
\begin{equation}
 \mathcal N=\Psi^{\mathsf T}R_f\Psi
 =\begin{pmatrix}\mathcal N_{11}&\mathcal N_{12}\\\mathcal N_{12}&\mathcal N_{22}\end{pmatrix}.
 \label{si-eq:gfN}
\end{equation}
Equation~(\ref{si-eq:gfN}) is a frame transformation, not an entrywise
identity: $\Psi=V_0^{-1/2}O$ for some orthogonal $O$, so $\mathcal N$ and
$V_0^{-1/2}R_fV_0^{-1/2}$ are orthogonally similar and share eigenvalues
and spectral norm, while their entries differ by that rotation. It is the
modal-frame entries that Eq.~(\ref{si-eq:gfabg}) uses.
So $\mathcal N_{11}$ and $\mathcal N_{22}$ are the in-span part of the
normalized resistance and $\mathcal N_{12}$ its transverse part, in the exact
sense of Eq.~(\ref{si-eq:gfspan}); $\mathcal N\succ0$, so
$\mathcal N_{11},\mathcal N_{22}>0$. Two
expansions now follow. First $f_{22}=4\mu_ss\mathcal N_{22}$ with
$\mathcal N_{22}=O(\epsilon_f)$, so $\nu=\mathcal N_{12}[1+O(\epsilon_f)]$ and
Eq.~(\ref{si-eq:gfexact}) gives
$\min_{\mu_s'>0}(r_+-r_-)/r_+=2\mathcal N_{12}[1+O(\epsilon_f)]$. Second
$\mu_s'^\star=c\mu_s[1+O(\epsilon_f)]$ by
Eq.~(\ref{si-eq:gfminimizer}), so the coupling strength evaluated at the
minimizer satisfies
$\epsilon_f=\lambda_{\max}(\mathcal N)[1+O(\epsilon_f)]$. Dividing the first
by the second proves the asymptote of Eq.~(\ref{eq:gapfloor}) with
$c_\star$ as in Eq.~(\ref{eq:gapfloorconstant}), the second form
following from
$\mathcal N_{12}=\tfrac12(\lambda_{\max}-\lambda_{\min})\sin2\varphi$
with
$\varphi=\tfrac12\operatorname{atan2}(2\mathcal N_{12},\mathcal N_{11}-\mathcal N_{22})$,
the quadrant-consistent convention the signed identity requires.

In particular $c_\star=0$ if and only if $\mathcal N_{12}=0$, that is if
and only if $R_f$ lies in the span or $\mathcal N$ is isotropic, and
$c_\star<1$ strictly because $\lambda_{\min}(\mathcal N)>0$. Explicitly, with
$a=1$,
$\eta_{\rm out}=1$ and $\lambda=\eta_{\rm in}/\eta_{\rm out}$,
\begin{equation}
 \mathcal N=\frac{1}{4\,\mathrm{Bq}_s}\widehat{\mathcal N},\qquad
 \widehat{\mathcal N}=\begin{pmatrix}
  \rho_{rr}/c&\hat\delta/\sqrt{cs}\\
  \hat\delta/\sqrt{cs}&\hat f_{22}/s
 \end{pmatrix},
 \label{si-eq:gfNhat}
\end{equation}
with $\mathrm{Bq}_s=\mu_s/(a\eta_{\rm out})$,
$\hat\delta=\Lambda\rho_{rr}+2\rho_{rp}$ and
$\hat f_{22}=\Lambda^2\rho_{rr}+4\Lambda\rho_{rp}+4\rho_{pp}$
evaluated at $\eta_{\rm out}=1$. The prefactor cancels in
Eq.~(\ref{eq:gapfloorconstant}), so
\begin{equation}
 c_\star=\frac{2\hat\delta}{\sqrt{cs}\;\lambda_{\max}(\widehat{\mathcal N})}
 =c_\star(\ell,c,\lambda)
 \label{si-eq:gfcstar}
\end{equation}
depends only on the degree, the area parameter combination and the
viscosity ratio, and is independent of $G_s$, of $a$, of $\mu_s$ and
of the overall fluid scale. The absolute floor itself carries the
expected surface-to-bulk scaling,
$\min_{\mu_s'>0}(r_+-r_-)/r_+=\hat\delta/(2\sqrt{cs}\,\mathrm{Bq}_s)
[1+O(\epsilon_f)]$. For the reference family of
Sec.~\ref{sec:couplingtradeoff} one has $\ell=2$, $\lambda=1$,
$c=5$, so $\Lambda=6$, $s=24$, $\hat\delta=45$,
$\lambda_{\max}(\widehat{\mathcal N})=\tfrac{65}{8}+\tfrac18\sqrt{3105}$ and
\begin{equation}
 c_\star=\frac{90/\sqrt{120}}
              {\tfrac{65}{8}+\tfrac18\sqrt{3105}}=0.544444,
 \label{si-eq:gfreference}
\end{equation}
against the computed $0.5443$ at $\epsilon_f=2\times10^{-4}$. Over
the surveyed grid Eq.~(\ref{si-eq:gfcstar}) gives
$c_\star\in[0.3615,0.9909]$ with median $0.7067$, the extremes at
$(\ell,c)=(2,11)$ and $(10,1)$, which reproduces the range computed at
finite coupling. Across the sampled grid, the constant increases with $\ell$,
decreases with $c$, and varies weakly with $\lambda$.

The floor describes a fixed harmonic at fixed $\ell,c,\mu_s$ and fluid
coefficients. The dependence on the shear viscosity is explicit:
$\mathcal N_{12}$ and $\mathcal N_{22}$ are both proportional to
$1/\mu_s$, so the absolute minimum gap tends
to zero as $\mu_s\to\infty$, but so does $\epsilon_f$ at the same
rate, and the invariant statement is the ratio. Minimizing the raw gap
over both surface viscosities without holding $\epsilon_f$ fixed gives
infimum zero, approached only along paths where the coupling itself
vanishes. Next, $c_\star$ is the leading constant for this one
objective and this one-parameter family. Modal independence controls
the membrane-only inverse with fixed modal patterns. In the coupled
problem the patterns depend on the material parameters, and their
derivatives enter the full-propagator material Jacobian of
Sec.~\ref{sec:coupledcontrol}. Its amplitude-projected smallest
singular value, rather than the rate gap or modal visibility alone,
controls local material sensitivity. The minimization concerns one harmonic. For an assembled band the
pencil is block diagonal in the membrane-only case and each degree
contributes its own transverse defect, so the per-degree formula bounds
the band degree by degree. A single $\mu_s'$ shared across degrees
requires a coupled minimization across the band.

\section{Observation designs used in the numerical verification}
\label{si-app:designs}
The designs below specify the schedules, weights, parameters and finite
grids behind the quantitative checks. Time means $t/t_0$ for a fixed
reference time $t_0$, and rates mean $r t_0$. Changing a mean rate does
not rescale the schedule. Moduli, surface viscosities and displacements
use fixed units $G_0$, $G_0t_0$ and $u_0$. Covariances are prescribed
numerical observation models in these displacement units. A single
record has $n$ entries; a pair has $2n$ entries in release-major order.

\subsection{Schedules, paired records and single-record conditioning}
The recurring schedules are
\begin{align}
 \mathcal T_1&=(0,\,0.2,\,0.4,\,0.6),\nonumber\\
 \mathcal T_2&=(0.05,\,0.11,\,0.23,\,0.42,\,0.7,\,1.05,\,1.6,\,2.4),
 \nonumber\\
 \mathcal T_3&=(0.35,\,0.52,\,0.79,\,1.18,\,1.74,\,2.8),\nonumber\\
 \mathcal T_4&=(0,\,0.07,\,0.23,\,0.7,\,1.5,\,3).
 \label{si-eq:schedules}
\end{align}
Zero-time samples have protocol-specific roles. The radial member of
a calibrated pair measures its common amplitude at zero; a generic
known-pattern record measures $S(P+Q)$. The equal-and-opposite record
has an identically zero mean and derivative row there, and its inverse
uses three strictly positive times. Both $\mathcal T_1$ and
$\mathcal T_4$ contain enough positive times for these comparisons.
For a paired record, let $K=(\rho^{|i-j|})_{i,j=0}^{n-1}$,
$v_i=-0.7+1.8i/(2n-1)$ and
$D_{ii}=0.8+0.4i/(2n-1)$, with $i=0,\ldots,2n-1$.
The three paired covariances are
\begin{align}
 \Sigma_{\rm I}&=I_{2n},\nonumber\\
 \Sigma_{\rm AR}&=D[\diag(1,1.44)\otimes K]D,\nonumber\\
 \Sigma_{\rm X}&=D\left[
 \begin{pmatrix}1&0.38\\0.38&1.44\end{pmatrix}\otimes K
 +\frac{0.12}{2n}vv^{\mathsf T}\right]D,
 \label{si-eq:weightfamilies}
\end{align}
with $\rho=0.55$ and $W=\Sigma^{-1}=L^{\mathsf T}L$.
The single-record comparison uses the first-record marginal covariance.
The paired figure uses $\mathcal T_2$ and $\Sigma_{\rm X}$.

At the figure's reference design, replacing the raw covariance by
$\Sigma_{\rm I}$ preserves the ordering of the calibrated pair,
independent-amplitude pair and single unknown-amplitude record over
the stated gap grid in both plotted coordinates. An equal-count iid
control takes two independent observations at each single-record time.
Its projected rate Gram matrix doubles, so the weak singular value
is multiplied by $\sqrt{2}$ while its quadratic gap power is unchanged.
This control uses independent repeated observations; its covariance
differs from $\Sigma_{\rm X}$.

The paired sensitivity sweep fixes $G_s=1$, $C_a=2$, $\ell=2$ and
$E_0=0.01$ in units $G_0u_0^2$. Put
$\chi=c+1-2/\Lambda$ and choose
\begin{equation}
 C=\sqrt{\frac{E_0}{2G_s(\chi\kappa^2+c)}},\qquad
 B_p=\kappa C,\qquad
 B_{\rm single}=\operatorname{sign}(\kappa)
             \sqrt{\frac{E_0}{2G_s\chi}} .
 \label{si-eq:designenergy}
\end{equation}
This fixes the true initial energies, while the inverse fits its
amplitude parameters. The comparison matches energy between one
and two acquisitions with their respective sample counts.
The $27$ designs are $\mathcal T_{1,2,3}$, the three covariances and
\begin{equation}
 (m,\kappa)\in\{(1.9,1.3),(0.8,-0.7),(3,2)\}.
 \label{si-eq:pairedtruths}
\end{equation}
At each design $r_T=m-d$, $r_D=m+d$, with $33$ half-gaps
$d_j=10^{-1-j/4}$, $j=0,\ldots,32$. All rates are positive;
negative $\kappa$ specifies the signed release ratio.
The final nine gaps determine the fitted log--log slopes.
Complex-step derivatives use $d=0.1,10^{-4},10^{-9}$.
The ninety-digit comparisons use $(m,\kappa)=(1.9,1.3)$,
$d=0.031,10^{-9}$, and the three schedule/covariance pairs
$(\mathcal T_1,\Sigma_{\rm I})$,
$(\mathcal T_2,\Sigma_{\rm X})$ and
$(\mathcal T_3,\Sigma_{\rm AR})$.

The $108$ unmixing and bound checks cross the same $27$ designs with
$d=0.1,10^{-5},10^{-9},0$. The $144$ labeled-inverse checks instead
cross the three schedules, the three $(m,\kappa)$ pairs,
$d=0.2,0.031,10^{-9},0$, both common-amplitude signs, and either the
two endpoint samples or every sample; $36$ are exact coincidences.
For the finite-calibration calculation, fix $\mathcal T_2$, $\Sigma_{\rm X}$,
$(m,\kappa)=(1.9,1.3)$ and the same $33$ gaps plus zero, with independent
ratio information $10^{-8},10^{-4},1$.
The energy-optimum checks cross $\mathcal T_{1,2,3}$ with
$r_T=r_D\in\{0.8,1.9,3\}$ and unit independent raw variances.
Each analytic optimum is compared with $200001$ uniformly spaced
fractions in $[10^{-7},1-10^{-7}]$.

The separate coincidence calculation fixes $(m,B)$, not energy:
\begin{equation}
 (m,B)\in\{(1.9,0.6),(0.8,1.1),(3,-0.4)\}.
 \label{si-eq:collisiontruths}
\end{equation}
It crosses $\mathcal T_1,\mathcal T_2$ with identity weight or
$\Sigma_{ij}=s_i s_j\,0.6^{|i-j|}$,
$s_i=0.8+0.4i/(n-1)$ for $i=0,\ldots,n-1$.
The $12$ designs each use $25$ half-gaps
$d_j=10^{-1-j/4}$, $j=0,\ldots,24$.
The coincidence figure uses $\mathcal T_2$, $W=I$, $m=1.9$, $B=0.6$,
and displays $d\ge10^{-5}$; its tangential initial energy is
$2G_s\chi B^2=4.08$. Its sensitivities scale linearly with $|B|$.
The finite alternative uses $d'=2d$, $B'=B/2$ before the stated
order-$d^2$ nuisance adjustment.

\subsection{Release-pair comparisons and coincidence brackets}
The reference mixed release pair is
$P=(0.83,0.41)^{\mathsf T}$,
$Q=(-0.29,0.66)^{\mathsf T}$, so $\det\mathcal M=0.6667$.
It uses $m=1.9$, common scale $S=7.3$, $\mathcal T_1$ and $W=I$.
Its full-gap sweep is
$r_T-r_D\in\{10^{-1},10^{-2},10^{-3},10^{-5},10^{-7},10^{-9},
10^{-12},0\}$; the convergence check also uses $10^{-6}$.
For the determinant sweep, keep the first row of $\mathcal M$ and
replace its second row by twice the first plus $\epsilon$ times the
reference second row, with
$\epsilon\in\{1,10^{-1},10^{-2},10^{-3},10^{-4},10^{-6},0\}$.
The reported ratio is $\sigma_{\min}/|\det\mathcal M|$.
The scale test uses $S=0.1,1,7.3,100$ at coincidence.

The four candidate poloidal releases are
\begin{equation}
 q=(0,1),\quad (1,0),\quad (\Lambda/2,1),\quad
 \left(\frac{\Lambda(\Lambda-2)+c\Lambda^2}{2c\Lambda},1\right).
 \label{si-eq:candidates}
\end{equation}
All six pairs are evaluated at coincidence with $W=I$ and $S=7.3$.
For the equal-coordinate-amplitude comparison, replace each $q$ by
$q/\|q\|_2$; for equal initial elastic energy, replace it by
$q/\sqrt{q^{\mathsf T}Kq/2}$ before forming the common-scale
nuisance projection. The $144$ designs are the Cartesian product
\begin{equation}
 \begin{gathered}
 \ell\in\{2,3,4,6\},\quad c\in\{0.5,1,5,20\},\quad
 r_T=r_D\in\{0.4,1.9,5\},\\
 t\in\{\mathcal T_1,\ (0,0.5,1),\ (0.1,0.3,0.9,1.5,2)\}.
 \end{gathered}
 \label{si-eq:releasegrid}
\end{equation}
This is a positive-stiffness algebraic grid. Its $36$ designs at
$c=0.5$ correspond to $C_a=-0.25$ and lie outside the usual
$C_a\ge0$ material range. The reference comparison uses
$\ell=2,c=5,r=1.9$ and $\mathcal T_1$.
Common-release homogeneity is checked at multipliers $2,7,100$.

The $400$ random coincidence designs use NumPy's PCG64 generator
with seed $20260914$. For each design, draw in this order:
two independent standard normal entries for $P$, two for $Q$,
$S=|N(0,1)|+0.1$, $r=|N(0,1)|+0.2$, and five independent
$|N(0,1)|$ values sorted into the common time schedule. Use $W=I$
and sixty-digit arithmetic for the projection and bracket.
The bracket tolerance is $10^{-40}$; lower and upper slack are the
minimum ratio to the lower bound and maximum ratio to the upper
bound over these designs.
For the unequal-schedule control the reference pair uses
$\mathcal T_1$ in its first record and $(0.05,0.3,0.55,0.9)$ in its
second. The factorized-weight control uses
$W_r=\bigl(\begin{smallmatrix}2&0.7\\0.7&1.3\end{smallmatrix}\bigr)$
and $W_t=\diag(0.3,1.7,0.9,2.4)$.
The nonfactorized control takes the next $8\times8$ normal draw $Z$
after the $400$ designs and uses $W=ZZ^{\mathsf T}+8I$.

\subsection{Pattern-order designs and finite root searches}
The nine pattern-order designs use $S=0.6$, $r=1.9$ and
$120$ decimal digits. Three have $L=I$ on $\mathcal T_2$ and
$(P,Q)=(1.3,0.4),(1,-3),(1.3,-1.3)$. Four more use the first
and third of these patterns on $\mathcal T_2$, with either
\begin{equation}
 L=\operatorname{chol}_{\rm lower}(ZZ^{\mathsf T}+8I)
 \quad\hbox{or}\quad
 L=\diag(1,10^{-1},\ldots,10^{-7}).
 \label{si-eq:patternroots}
\end{equation}
Here $Z$ is the first $8\times8$ standard normal draw from a freshly
seeded NumPy PCG64 generator with seed $20260914$; its double-precision
Cholesky entries are then represented at the working precision.
In both cases the weight is $W=L^{\mathsf T}L$. The diagonal root has
seven decades of spread and the weight has fourteen.
The last two designs use $(P,Q)=(1.3,0.4),(1.3,-1.3)$,
$L=I$ and $(0.12,0.65,1.8)$.
Constants are evaluated at full gap $d=10^{-10}$ and exponents from
the ratio between $d=10^{-9}$ and $10^{-10}$.
The symbolic minor enumeration uses moment indices $0,\ldots,6$.

The finite multi-start search uses sixty working digits and
squared-residual tolerance $10^{-40}$ on the three equations
$S(Pe^{-r_Tt_i}+Qe^{-r_Dt_i})=y_i$.
Its $140$ initial triples are
\begin{equation}
 \begin{aligned}
 S&\in\{0.5,1,1.5,2\},\\
 r_T&\in\{0.5,1.2,2,3.5,6,9,14\},\\
 r_D&\in\{0.2,0.7,1.4,2.5,4\}.
 \end{aligned}
 \label{si-eq:startgrid}
\end{equation}
The four exact-data cases have true scale $S=1$ and the inputs
\begin{equation*}
 \begin{array}{c|c|c}
  (P,Q)&(t_1,t_2,t_3)&(r_T,r_D)\\ \hline
  (1,3.862)&(0.6728,2.1558,2.5032)&(2.981473,1.805891)\\
  (1,2)&(0.3,1,2.5)&(2.3,1.5)\\
  (1,-1.5)&(0.2,1.1,2.6)&(3.2,0.6)\\
  (1,-1)&(0.2,1.1,2.6)&(3.2,0.6).
 \end{array}
\end{equation*}
Accepted solutions satisfy $S\ne0$ and $r_T>r_D>0$.
Two solutions are identified when both rates differ by less than
$10^{-12}$. Unconverged starts are discarded.
The root finder is the multidimensional Newton iteration with numerical
Jacobian and line search (mpmath's default multidimensional solver).
The search finds two branches in each generic case and one in the
equal-and-opposite control. Global uniqueness in the latter follows
from the analytic argument of Sec.~\ref{sec:collision}.
The separate unique generic example uses $(P,Q)=(1,-2)$, $S=1$,
times $(1,2,3)$ and decay factors $(0.9,0.99)$; exact real-root
isolation of the eliminated quartic determines its admissible count.

\subsection{Gap-floor grids and coupled responses}
These calculations use the zero-bending, one-harmonic poloidal
problem. Here $\epsilon_f$ is evaluated on its $2\times2$ block,
and a gap-floor target is evaluated at the minimizing $\mu_s'$.
All gap-floor grids use $G_s=a=1$. The $135$ comparisons of the exact
minimum with numerical minimization cross
\begin{equation}
 \begin{gathered}
 \ell\in\{2,3,5,8,12\},\quad
 \lambda\in\{0.1,1,10\},\quad c\in\{1,5,11\},\\
 \mu_s\in\{0.5,1.9,40\},\qquad
 \eta_{\rm out}=10^{-3},\quad \eta_{\rm in}=\lambda\eta_{\rm out}.
 \end{gathered}
 \label{si-eq:135grid}
\end{equation}
Brent minimization acts on $\log\mu_s'$ with tolerance $10^{-15}$,
initialized by the bracket
\begin{equation*}
 \bigl(\log(0.7\mu_s'^\star),\ \log\mu_s'^\star,\
       \log(1.4\mu_s'^\star)\bigr).
\end{equation*}
Thus the numerical objective is evaluated independently, while the
analytic optimizer supplies its bracket.

The $1500$ interiority tests cross
$\ell\in\{2,4,6,10,14\}$,
$c\in\{0.2,1,5,11\}$,
$\mu_s\in\{10^{-3},1,10^3\}$ and, independently for each fluid,
$\eta_{\rm in},\eta_{\rm out}\in\{10^{-9},10^{-6},10^{-3},1,10^3\}$.
They test the strict inequality in Eq.~(\ref{si-eq:gfinterior}) at
unit radius. As with Eq.~(\ref{si-eq:releasegrid}), the $c<1$ subset is an
algebraic positive-stiffness stress test.
The separate two-viscosity grid is
$\ell\in\{2,5,9\}$, $c\in\{1,5\}$ and independent
$\mu_s,\mu_s'\in\{0.01,1,100\}$, at
$\eta_{\rm in}=\eta_{\rm out}=10^{-3}$.
Scale independence uses $\ell=2,c=5$,
$\mu_s\in\{10^{-2},1,10^2,10^4\}$,
$a\in\{1,3.5\times10^{-6}\}$ and equal fluid viscosities $10^{-4}$.

The $126$-design constant survey crosses
$\ell\in\{2,3,4,6,8,10\}$,
$\lambda\in\{0.1,0.2,0.5,1,2,5,10\}$ and $c\in\{1,5,11\}$,
with $\mu_s=1.9$. Both the closed form and numerical minima at
target $\epsilon_f=10^{-4},10^{-3}$ are evaluated.
The target is imposed at the minimizing $\mu_s'$, by adjusting
$\eta_{\rm out}$ at fixed $\lambda$.
The four asymptotic families use $(\ell,c,\lambda)=(2,5,1)$,
$(4,1,10)$, $(10,11,0.1)$ and $(3,5,0.5)$, with $\mu_s=1.9$
and targets $10^{-3},10^{-4},10^{-5}$.

The separated-rate coupled sweep fixes
$\ell=2,c=5,\mu_s=1.9,\mu_s'=1.3$ and
$\eta_{\rm in}=\eta_{\rm out}\in
\{0,10^{-4},10^{-3},10^{-2},10^{-1},1,10\}$.
For modal angles, normalize the two eigenvectors in Euclidean
$(q_r,q_p)$ coordinates and match them by sorted eigenrate to the
separated reference eigenvectors.
The coupled normal records start from $q_T=(0,1)$ and $q_D=(1,0)$.
For the reference-family gap-floor plot, use the same $\mu_s$ and
equal fluid viscosities $10^{-4},3\times10^{-4},10^{-3},
3\times10^{-3},10^{-2},3\times10^{-2},10^{-1}$, minimizing over
$\log\mu_s'\in[\log1,\log300]$ with tolerance $10^{-12}$.
Projected sensitivities there use $\mathcal T_1$, $W=I$, common
scale one and raw viscosity coordinates.

For the invisible-mode/material-rank example, use
$\ell=2,c=5,G_s=a=\mu_s=\eta_{\rm in}=\eta_{\rm out}=1$,
$\mu_s'=55/8$, the same $q_T,q_D$, $\mathcal T_1$, $W=I$ and
common scale one. Write $D=V+R_f$ and $A=D^{-1}K$.
The two material derivative columns are stacked normal components of
\begin{equation}
 \partial_{\mu_j}e^{-tA}q
 =\mathcal L_{\exp}(-tA,tD^{-1}R_jA)q,
 \qquad R_j\in\{R_s,R_d\},
 \label{si-eq:hiddenfrechet}
\end{equation}
where $\mathcal L_{\exp}(X,H)$ is the Fr\'echet derivative of the
matrix exponential at $X$ in direction $H$.
For stacked mean $y$, the amplitude projection is
$I-yy^{\mathsf T}/(y^{\mathsf T}y)$.
This tests material rank including the derivatives of modal amplitudes.

\subsection{Coupled rank certificates and their aggregation}
The $48$ designs at each nominal coupling use one poloidal harmonic
and cross
$\ell\in\{2,3,4,6\}$,
$\lambda\in\{0.1,1,5,10\}$ and reference half-gap
$d\in\{0,0.05,0.4\}$. Set
$G_s=a=1$, $c=5$, $m=1.3$, $C=0.05$, $\kappa=-1.4$ and
$t=\mathcal T_4$.
Then $\mu_s=G_s/(m-d)$, $\mu_s'=G_sc/(m+d)$ and
$q_T=(0,2\kappa C/\Lambda)$, $q_D=(C,0)$.
At each point scale the fluid block with viscosity ratio $\lambda$
once to obtain $\epsilon_f\in\{0,10^{-3},10^{-2},10^{-1},1\}$,
and hold that physical block fixed during differentiation.
The grid contains $192$ positive-coupling cases and $48$
membrane-only baselines.
The raw covariance and weight are
\begin{equation}
 \Sigma=\begin{pmatrix}1&0.25\\0.25&1.4\end{pmatrix}\otimes T,
 \qquad T_{ij}=0.4^{|i-j|},\qquad W=\Sigma^{-1}=L^{\mathsf T}L .
 \label{si-eq:coupledcovariance}
\end{equation}

The certificate count uses conservative scalar bounds, not the computed
Jacobian difference. With the notation of
Eqs.~(\ref{eq:fluidforward}) and (\ref{eq:fluidderivative}), define
\begin{align}
 a_j&=\|e_r^{\mathsf T}K^{-1/2}\|_2\,\|K^{1/2}q_j\|_2,\nonumber\\
 p_i&=\delta_f b t_i e^{-\alpha t_i},\nonumber\\
 h_{ik}&=\delta_f b g_k(2t_i+bt_i^2)e^{-\alpha t_i},\nonumber\\
 \overline\eta&=\frac{\|L\|_2}{|C|}
       \left(\sum_{i,j}a_j^2p_i^2\right)^{1/2},\nonumber\\
 \overline\zeta&=\|L\|_2
       \left(\sum_{i,j,k}a_j^2h_{ik}^2\right)^{1/2}.
 \label{si-eq:boundaggregation}
\end{align}
Here $j$ indexes the two releases and $k$ the two log-viscosities.
Let $J_0$ and $n_0$ be the whitened reference material and common-scale
derivative blocks, and $s_0=\sigma_{\min}(P_{n_0}^{\perp}J_0)$.
Count a design only when
\begin{equation}
 \overline\eta<\|n_0\|_2,\qquad
 s_0-\overline\zeta
 -\frac{\overline\eta}{\|n_0\|_2}\|J_0\|_2>0.
 \label{si-eq:countcriterion}
\end{equation}
Thus the scalar bounds are combined by a Frobenius norm and then
multiplied by $\|L\|_2$. This prescription yields the stated
$20$ of $48$ at $\epsilon_f=0.01$, including $12$ of the $16$
exact reference coincidences. A nonpositive lower bound is inconclusive.
Complex-step columns use step $10^{-20}$; the independent five-point
difference at $\epsilon_f=0.1$ uses log-viscosity step $10^{-4}$.
The finite-calibration checks use information $0.01,1,100$.

The $96$ independent traction checks cross
$\ell\in\{2,3,4,6\}$, $a\in\{0.7,2.3\}$,
$(\eta_{\rm in},\eta_{\rm out})\in\{(1,0),(0,1),(5,1)\}$
and boundary velocities $(1,0),(0,1),(0.4,-0.7),(1,1)$.
Rigid-drag controls use $\ell=1,a=1.7,\eta_{\rm in}=2.1$,
$\eta_{\rm out}=0.3$.

\subsection{Additional continuum and exponential-sum checks}
The diagonal membrane integration uses
$(G_s,C_a,\mu_s,\mu_s')=(1,2,0.7,2.1)$,
initial modal amplitudes $(0.8,0.6)$ and output times
$(0.1,0.2,0.4,0.8,1.2)$. DOP853 integration on $[0,1.2]$
uses relative tolerance $2\times10^{-12}$ and absolute tolerance
$2\times10^{-14}$. Strain quadrature uses $\ell=2,3,4$,
Gauss--Legendre orders $4,8,16,32$ and composite Simpson counts
$32,64,128,256$. The signed four-time shape recovery uses
$\mathcal T_1$, $(r_T,r_D)=(10/7,50/21)$ and $B=0.6$.
The independent matrix solver uses $\ell=2,3,4,6$,
$32$-point Gauss--Legendre assembly,
$(G_s,C_a,\mu_s,\mu_s')=(1,2,0.7,2.1),(2,0.3,1.3,0.8),
(1,2,0.7,3.5)$, initial tangential amplitude $10^{-3}$,
final time $1.2$ and $40,80,160,320$ fixed RK4 steps.

The $36$ three-positive-time inverse cases use schedules
$(0.12,0.65,1.8)$, $(0.3,0.7,2.7)$ and $(0.02,0.11,0.47)$.
Python's Mersenne Twister, seeded with $20260912$, supplies $12$
draws per schedule in this order:
$\log m$ uniform on $[\log0.45,\log3]$,
$d/m$ uniform on $[0.08,0.88]$, and
$\log|B|$ uniform on $[\log0.15,\log2]$.
The sign of $B$ alternates, starting positive.
The added small-gap case is $(B,m,d)=(-0.7,1.3,10^{-4})$
on the first schedule.

For the $100$-digit matrix comparison, the single-record schedules are
$(0.1,0.4,0.9,1.8)$, $(0.2,0.31,0.55,0.94,1.6,2.3)$ and
$\mathcal T_2$.
The four $(m,A,C,B)$ tuples are
$(0.8,0.7,-1.2,0.6)$, $(1.9,1.3,-1.3,-0.7)$,
$(3,-2,0.4,1.1)$ and $(1.4,0.9,0.9,-1.2)$.
Each uses half-gaps $10^{-j}$, $j=1,\ldots,8$, and three covariances:
identity, $K_{ij}=0.55^{|i-j|}$, and $D K D$ with
$D_{ii}=0.8+0.4i/(n-1)$. This gives $36$ designs.
The $12$ paired designs cross $(0.2,1.4)$ and the six-time schedule
above with the same three covariance constructions on all $2n$
release-major entries, and
$(m,B_p,C)=(1.9,1.3,0.7),(0.8,-0.7,1.1)$.
The paired relative ratio is $B_p/C$.
Direct singular-value decomposition is checked against a Gram
eigensolve; the final decade determines the reported powers.
\section{Detailed numerical verification}
\label{si-app:verificationdetails}
\subsection{Four-sample exponential reconstruction}\label{si-app:prony}
Spatially labeled projections are sufficient but not necessary
to resolve two distinct rates. A scalar mixture sampled at
$t_n=n\Delta t$ has the form
\begin{equation}
 y_n=A_Tx_T^n+A_Dx_D^n,\qquad
 x_i=e^{-r_i\Delta t},\quad n=0,1,2,3 .
 \label{si-eq:prony}
\end{equation}
For nonzero $A_T,A_D$, $\Delta t>0$, and $x_T\ne x_D$, define
\begin{equation}
 \calD=y_0y_2-y_1^2=A_TA_D(x_T-x_D)^2\ne0.
 \label{si-eq:hankel}
\end{equation}
The recurrence
\begin{equation}
 y_2=\hat s\,y_1-\hat p\,y_0,\qquad y_3=\hat s\,y_2-\hat p\,y_1
 \label{si-eq:rec}
\end{equation}
has determinant $\calD$, giving $\hat s=x_T+x_D$ and $\hat p=x_Tx_D$.
The roots of $x^2-\hat sx+\hat p$ recover the unordered pair of rates.
This is the two-exponential Prony construction cited in the article. A mode label or a separately justified ordering
is needed to associate each recovered rate with a particular
surface viscosity. Local reconstruction around a distinct labeled
pair is regular, while global exchange of the modes remains an
ambiguity of the scalar record.

At coincident rates, $\calD=0$ and the time history becomes a
single exponential, possibly identically zero when the signed
amplitudes cancel. The two rate-derivative columns are then
proportional even after projecting the unknown amplitudes.
In contrast, separately labeled observations
in Eq.~(\ref{eq:observe}) retain Eq.~(\ref{eq:modalinverse}) at
$r_T=r_D$. The channel count is therefore determined by the
response and observation together, rather than by the number of sensors.

\subsection{Verification of the specified models}
The verification programs check spherical strain forms,
coupled weak-matrix time integration, the prescribed shape
responses, and nuisance-projected sensitivities.
Positive-definite correlated numerical covariances test the
effect of transforming and weighting the complete observation.
Constructed-drag regression tests are kept
separate from the membrane model.

For the continuum example we use nondimensional material values
$G_s/G_0=1$, $C_a=2$,
$\mu_s/(G_0t_0)=0.7$, and $\mu_s'/(G_0t_0)=2.1$,
with arbitrary fixed reference modulus $G_0$ and time $t_0$.
Consequently $r_Tt_0=10/7$ and $r_Dt_0=50/21$.
Displacement amplitudes are normalized by a common small
reference amplitude, and plotted time denotes $t/t_0$.
The preselected harmonic degrees
$\ell=2,3,4$ are checked by angular quadrature against the
exact strain norms. These forms provide analytic
benchmarks, while numerical integration checks the corresponding
amplitude histories. The scalar mixed-mode reconstruction tests
Eq.~(\ref{si-eq:hankel}), both distinct-rate recovery and the
coincident-rate degeneracy. The programs record
resolution choices, tolerances, and actual errors. These checks
verify the specified mathematical implementations. Representative
values are collected in Table~\ref{tab:verification}.

\begin{figure*}[t]
 \includegraphics[width=\textwidth]{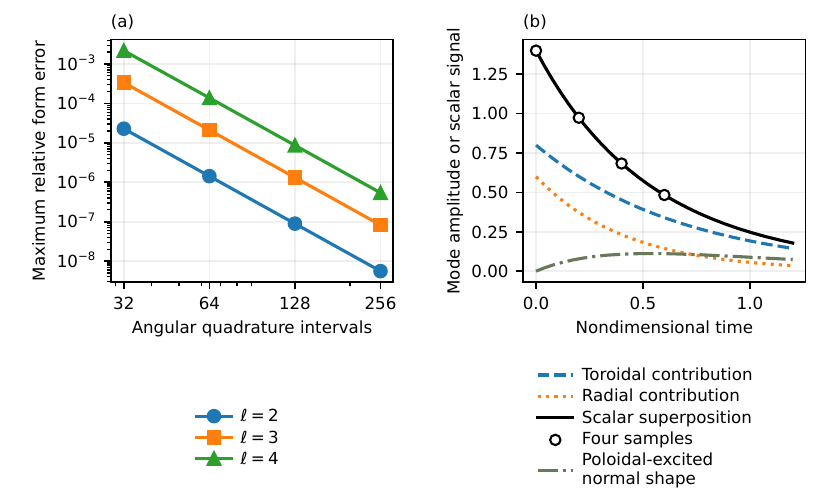}
 \caption{Linear-membrane quadrature and relaxation in the reference
 model. (a) Composite Simpson convergence of the largest relative
 error in the shear and dilatational forms, for prescribed harmonic
 degrees $\ell=2,3,4$. (b) Toroidal and radial amplitudes
 $0.8e^{-(10/7)t}$ and $0.6e^{-(50/21)t}$, their scalar sum, and
 four equally spaced samples. The scalar sum illustrates the Prony
 reconstruction; its toroidal contribution is absent from normal imaging.
 The dash-dotted curve is the normal-shape
 signal from poloidal excitation in Eq.~(\ref{eq:poloidal}), with
 $\Lambda A/2=0.6$. Time and amplitudes use the reference units
 specified in Sec.~\ref{sec:verification}, which also describes the
 verification.}
 \label{si-fig:modes}
\end{figure*}

Figure~\ref{si-fig:modes} shows the quadrature convergence and the
reference relaxation histories.

For $\ell=4$, four-node Gauss--Legendre quadrature
underintegrates the forms; eight or more nodes recover them
to the accuracy in Table~\ref{tab:verification}.
Table~\ref{si-tab:independent} collects the remaining verifications of
Secs.~\ref{sec:paired} to \ref{sec:fluid}, with their coverage
and largest deviations.

The second membrane implementation is independent of the first. It
assembles the poloidal and radial weak matrices and integrates their
coupled coordinate dynamics with classical fourth-order Runge--Kutta
stepping, so agreement with the modal solution tests the eigenfamily
construction of Sec.~\ref{sec:eigenfamilies} rather than repeating it.

Hyperbolic series and QR elimination resolve the small projected
derivatives, with a complex-step differentiation check. Direct
double-precision subtraction is retained as a roundoff control.
Exact coincidence and separately labeled modal records are evaluated
directly.

The small-gap case in Table~\ref{si-tab:independent} illustrates how
coincidence sensitivity, rather than the fit residual, limits rate
recovery: the forward fit remains accurate as the inferred gap loses
relative accuracy.

The energy-matched paired comparison uses
Eq.~(\ref{si-eq:designenergy}) for the simulated states and fits their
amplitudes. Unequal variances and cross-record correlation are included
through Eq.~(\ref{si-eq:weightfamilies}). Both labeled-rate and
log-viscosity coordinates recover the predicted sensitivity powers
zero, one and two for calibrated paired, uncalibrated paired and
single-record data.

\subsection{Matrix comparison and bulk-resistance checks}
Separate code paths compare a direct singular-value decomposition
with a Gram eigensolve of the same derivative matrix. This checks
the linear algebra on shared derivative inputs. Analytic derivatives
are also checked by arbitrary-precision numerical differentiation
and by central differences. The high-precision calculations use
60--120 decimal digits, with the precision and designs specified
in Sec.~\ref{si-app:designs}.

The bulk-resistance verification reconstructs traction vectors from radial
Stokes powers independently of the closed-form matrix entries of
Eq.~(\ref{eq:fluidmatrix}). Rigid translation and rotation at $\ell=1$
give additional drag checks, separately from the admissible $\ell\ge2$
elastic dynamics.

The coupled grid tests the perturbation statements with the physical
bulk resistance held fixed during material differentiation.
The projected criterion certifies all $48$ designs at
$\epsilon_f=10^{-3}$ and $20$ at $10^{-2}$, including $12$ of the
$16$ reference coincidences at the latter coupling.
Its lower bound is inconclusive at $10^{-1}$ and $1$.
The covariance transformation and finite-calibration Schur bounds
are checked using the coupled means and derivatives.

\begin{table*}[t]
\caption{Numerical verifications of Secs.~\ref{sec:paired} to
\ref{sec:fluid}. Deviations are relative unless stated, are the
largest observed over the stated coverage, and are rounded up to two
significant figures; they are not error estimates.}
\label{si-tab:independent}
\begin{ruledtabular}
\footnotesize
\begin{tabular}{p{0.24\textwidth}p{0.40\textwidth}p{0.28\textwidth}}
Verification & Coverage & Largest deviation\\
\hline
Second membrane solver &
 $\ell=2,3,4,6$, three material choices including equal rates,
 refinement from 40 to 320 steps &
 observed order at least $4.01$; normal response at 320 steps
 $4.6\times10^{-10}$\\
Finite-time conditioning &
 two schedules, two weights and three $(m,B)$ choices, 12 cases with 25
 half-gaps each,
 $10^{-1}$ to $10^{-7}$ &
 complex-step derivative $2.3\times10^{-16}$ absolute\\
Three-time inverse &
 36 seeded choices, three unequal-time schedules, both amplitude signs &
 parameters $6.7\times10^{-12}$; forward residual
 $3.9\times10^{-15}$\\
\quad at the added half-gap $d=10^{-4}$ &
 the same inverse at one smaller gap &
 gap $4.0\times10^{-7}$; forward residual $3.1\times10^{-15}$\\
Paired protocol &
 three schedules, three covariance designs, three mean-rate and ratio
 choices: 27 designs with 33 half-gaps each, $10^{-1}$ to $10^{-9}$,
 891 positive-gap rows &
 smallest-sensitivity powers $0$, $1$, $2$ recovered\\
Double precision against 90-digit arithmetic &
 six schedule/covariance and gap combinations specified in
 Sec.~\ref{si-app:designs}, both coordinate sets; complex step at
 all 27 paired designs &
 agreement $3.2\times10^{-15}$; complex step $1.2\times10^{-16}$
 absolute\\
Full-covariance unmixing and explicit bound &
 108 cases &
 rate-Gram difference $1.6\times10^{-15}$\\
Labeled inverse &
 144 tests including 36 exact coincidences &
 recovery $5.9\times10^{-16}$\\
Finite calibration information &
 $\mathcal I_\kappa=10^{-8},10^{-4},1$ at 33 positive half-gaps and at
 exact coincidence &
 Schur bound of Eq.~(\ref{eq:pairedcalibration}) holds throughout\\
Equal-variance energy optimum &
 nine schedules and rates against 200001-point fraction grids &
 closed-form against grid optimum fraction $2.3\times10^{-6}$ absolute;
 at the eight-time schedule with $r_Tt_0=r_Dt_0=1.9$, optimum
 tangential energy fraction $0.3300183971$ and
 $\kappa_*^2=0.4346279093$\\
Matrix comparison at 100 digits &
 36 single-record and 12 paired designs, fixed positive-time schedules,
 three covariance choices, eight half-gaps from $10^{-1}$ to
 $10^{-8}$ &
 decomposition against Gram eigensolve $9.7\times10^{-50}$; asymptotic
 constant $2.9\times10^{-8}$; final-decade power $1.2\times10^{-7}$\\
Bulk resistance &
 96 traction vectors reconstructed from radial Stokes powers, with
 rigid translation and rotation at $\ell=1$ &
 $1.5\times10^{-15}$\\
Coupled grid &
 $\ell=2,3,4,6$; $\lambda=0.1,1,5,10$; reference half-gaps $0,0.05,0.4$;
 nominal $\epsilon_f=0,10^{-3},10^{-2},10^{-1},1$:
 192 coupled cases and 48 membrane-only baselines, giving
 480 sorted-rate inequalities, 1440 propagator bounds and 2880
 material-derivative bounds &
 all bounds hold; Fr\'echet against complex step
 $7.9\times10^{-15}$, against five-point differences at
 $\epsilon_f=10^{-1}$, $1.5\times10^{-11}$\\
Projected criterion of
 Eq.~(\ref{eq:fluidprojected}) &
 48 designs at each coupling, including 16 exact reference-rate
 coincidences &
 positive in 48 of 48 at $\epsilon_f=10^{-3}$ and 20 of 48 at
 $10^{-2}$; inconclusive at $10^{-1}$ and $1$
\end{tabular}
\end{ruledtabular}
\end{table*}

\subsection{Deterministic coupled-model inversion}
\label{si-app:modelmismatch}
Paired records generated with bulk dissipation permit a direct comparison
of matched and misspecified fits. We select $\ell=2$, $\lambda=1$,
$G_s=a=1$, $c=5$, $m=1.3$, $d=0.05$, $C=0.05$ and $\kappa=-1.4$,
with schedule $\mathcal T_4$ and the weight in
Eq.~(\ref{si-eq:coupledcovariance}).
Adaptive DOP853 integration generates the deterministic records at
$\epsilon_f=0,10^{-3},10^{-2},0.1,0.42,1$, evaluated at the prescribed
material pair. Matrix-exponential fits use Fr\'echet derivatives and
optimize $\log\mu_s$, $\log\mu_s'$ and $\log C$, holding the physical
bulk resistance fixed in the coupled model and setting it to zero
in the membrane-only model.
At each coupling, six initial guesses distinct from the prescribed
parameters converge to matching estimates for both models.
The coupled fits recover both viscosities to numerical accuracy;
the two models agree at $\epsilon_f=0$.

Table~\ref{si-tab:modelmismatch} reports the membrane-only fits for the four
larger couplings. The signed viscosity bias is
$100(\widehat\mu/\mu-1)$ percent, and the weighted relative residual is
$100\|L(\widehat y-y)\|_2/\|Ly\|_2$ percent.
In this example, small signal residuals coexist with substantial
viscosity bias when bulk dissipation is omitted.

\begin{table}[t]
\caption{Membrane-only fits to deterministic coupled records.
Biases and residuals are percentages, rounded to three significant
figures, for the fixed design of Sec.~\ref{si-app:modelmismatch}.}
\label{si-tab:modelmismatch}
\begin{ruledtabular}
\begin{tabular}{rrrr}
$\epsilon_f$ & $\mu_s$ bias & $\mu_s'$ bias & Weighted residual\\
\hline
$0.01$ & $+1.00$ & $+0.944$ & $0.0171$\\
$0.1$  & $+10.0$ & $+9.41$  & $0.172$\\
$0.42$ & $+41.9$ & $+39.2$  & $0.717$\\
$1$    & $+99.6$ & $+92.7$  & $1.55$
\end{tabular}
\end{ruledtabular}
\end{table}

\subsection{Release design, coupling, and the two closed forms}
Further checks address the release geometry and the closed forms in
Secs.~\ref{sec:independence} and \ref{sec:couplingtradeoff}
of the article and Secs.~\ref{si-app:pattern} and \ref{si-app:gapfloor} here.

The release-design calculation evaluates coincidence directly and
checks the prescribed modal directions against assembled weak
matrices at $\ell=2,3,4,6$. It also checks the release-to-record map
and the holding traction. Translation overlaps of the axisymmetric
loading are evaluated with $64$-point Gauss--Legendre quadrature;
the other force and torque components vanish by harmonic
orthogonality. The mixed reference pair of Sec.~\ref{si-app:designs}
has coincidence margin $1.404$. Along the determinant sweep, its
margin approaches a linear dependence on $|\det\mathcal M|$.

For a common amplitude rescaling, every projected rate column scales
by that amplitude. Equal-energy comparisons therefore normalize each
release to its energy budget before forming the projected Jacobian.
The resulting spread across release pairs is reported below.

The coupling verification evaluates the membership defect of
Eq.~(\ref{eq:transversedefect}), which is $45$ at $\ell=2$ with equal
fluid viscosities, then the rotation, amplitude and cross-talk sweeps
quoted in Sec.~\ref{sec:couplingtradeoff}, then the gap floor over the
$126$-design grid. At the worst attainable case, evaluated with the
coupled parameter-dependent patterns at
$\epsilon_f=2.0\times10^{-4}$, the modal determinant is $-3.56$ and
the projected singular values are $0.0206$ and $0.486$.

Symbolic checks confirm the congruence identity, transverse defect
and discriminant splitting of Sec.~\ref{si-app:gapfloor}. The exact
minimum agrees with one-dimensional numerical minimization to
$1.4\times10^{-10}$ over the $135$ designs of
Eq.~(\ref{si-eq:135grid}). In four families, the minimum relative gap
divided by its coupling strength approaches the closed-form
$c_\star$ as $\epsilon_f$ decreases from $10^{-3}$ to $10^{-5}$.
The interiority condition fails in nine of the $1500$ extreme
designs, all with $\lambda_{\max}(\mathcal N)\ge8.8$;
all designs with $\lambda_{\max}(\mathcal N)\le1$ have an interior
minimum. The separate two-viscosity grid retains a positive relative
gap, with a sampled minimum $4.6\times10^{-5}$.

The pattern-order calculation checks Lemmas 1 and 4 symbolically.
The leading minors are $(0,1,2)$ at full-gap order $d$ for a generic
pattern and $(1,2,3)$ at order $d^4$ for the equal-and-opposite one.
The nine designs of Sec.~\ref{si-app:designs} reproduce the predicted
orders and constants across dense and diagonal weights and both
three-time and eight-time schedules. Rank-loss controls remove an
excited component, reduce the schedule to two times, or replace one
of three positive times by zero in the equal-and-opposite case.
They confirm the sampling distinction of Lemma 3. The coefficient
check also recovers the factor $3/2$ produced by freezing the
nuisance direction, as derived after Eq.~(\ref{si-eq:patBdcol}).
\subsection{Release survey and coupling-check details}
Over the algebraic $144$-design grid of Eq.~(\ref{si-eq:releasegrid}),
the median sensitivity spread across release pairs is $3.70$ at
equal coordinate amplitude and $2.49$ at equal initial energy.
The trace-free and normal-traction-only pair is the least sensitive
in $105$ of those designs at equal coordinate amplitude and $83$ at equal
energy. These counts describe the specified positive-stiffness grid,
including its $36$ algebraic controls with $c=0.5$. At the reference design $c=5,\ell=2$,
the corresponding spreads are $9.24$ and $3.88$.

For the separated-rate reference family of
Sec.~\ref{si-app:designs}, the largest modal-direction rotation is
about $9.1^\circ$ per unit $\epsilon_f$, giving $0.33^\circ$ at
$\epsilon_f=0.036$ and $3.3^\circ$ at $\epsilon_f=0.36$.
If $\mathcal M_0$ is the membrane-only modal-amplitude matrix and
$\mathcal M_f$ the coupled one for the same two releases, the
reported cross-talk is
$\|\mathcal M_0^{-1}\mathcal M_f-I\|_2$.
Its ratio to $\epsilon_f$ is $0.680$, $0.679$ and $0.670$ at
$\epsilon_f=3.6\times10^{-4}$, $3.6\times10^{-3}$ and
$3.6\times10^{-2}$, and $0.587$ at $\epsilon_f=0.36$.
Thus the small-coupling coefficient is about $0.68$ for this design.
These modal comparisons require a separated reference spectrum;
the coordinate-free propagator bounds retain their meaning at a
reference coincidence.

The reference-family asymptotic gap coefficient is
$c_\star=0.544444$ in closed form. The finite-coupling ratios are
$0.5443$, $0.5441$, $0.5432$ and $0.5409$ at
$\epsilon_f=2.0\times10^{-4}$, $6.0\times10^{-4}$,
$2.0\times10^{-3}$ and $6.0\times10^{-3}$.
Across the $126$ designs specified in Sec.~\ref{si-app:designs},
the closed-form coefficient ranges from $0.3615$ to $0.9909$,
with median $0.7067$ and $34\%$ of designs below the reference
value. The ratios at the two finite target couplings differ by at
most $1.1\times10^{-3}$ relatively over that grid.
The coefficient is specific to the degree, elastic area parameter and
fluid viscosity ratio.

\section{Spherical strain norms}\label{si-app:norms}
On the unit sphere, the toroidal vector harmonic satisfies
$\nabla_\Omega\cdot\phi_T=0$ and
$\int|\phi_T|^2\,d\Omega=\Lambda$.
Integration by parts with the unit-sphere Ricci curvature gives
\begin{equation}
 2\int|\operatorname{sym}\nabla_\Omega\phi_T|^2\,d\Omega
   =(\Lambda-2)\int|\phi_T|^2\,d\Omega .
\end{equation}
Factors of $a^{-1}$ from surface derivatives cancel the
$a^2$ area factor, giving the toroidal entry of
Eq.~(\ref{eq:toroidalnorm}). For the radial field,
$\tr E=2Y_{\ell m}/a$, giving $R_d=4$.
The mean-zero and rigid-orthogonality properties follow from
harmonic orthogonality for $\ell\ge2$. Because the vanishing
trace or deviatoric strain is pointwise, the eigenfamily proof
holds against all admissible test fields.

\section{An explicit finite-gap alternative}
\label{si-app:alternative}
Using the notation of Sec.~\ref{sec:collision}, choose the unique
real coefficients $\alpha_0,\beta_0$ satisfying
\begin{equation}
 \alpha_0x_1-\beta_0x_2
   =-\frac{k^2-1}{6}P_{[x_1,x_2]}x_3 ,
\end{equation}
where $P_{[x_1,x_2]}$ projects onto the indicated span.
Its Gram matrix is nonsingular at distinct positive sample
times. Set
\begin{equation}
 B'=\frac{B}{k}(1+\alpha_0d^2),\qquad
 m'=m+\beta_0d^2,\qquad d'=kd .
\end{equation}
All coefficients are fixed as $d\to0$.
Expansion of the alternative signal gives
\begin{align}
 L(y'-y)
 &=2Bd^3\left[\alpha_0x_1-\beta_0x_2+
                    \frac{k^2-1}{6}x_3\right]+O(d^5)\\
 &=\frac{B(k^2-1)}{3}d^3b_3+O(d^5).
\end{align}
For sufficiently small $d$, $m'>kd>0$,
$B'$ has the sign of $B$, and $|B'|<|B|$ for $k>1$.
Both excitations therefore satisfy the same upper amplitude
bound. This explicit construction establishes the weighted data distance in
Eq.~(\ref{eq:finitedistance}) for the stated finite-gap alternative
under the fixed weight and amplitude bound.

\section{Joint and amplitude-projected coincidence expansions}
\label{si-app:comparison}
Here $d$ is the article's half-gap and $x_k=L(t_j^ke^{-mt_j})_j$.
The projected block $M$ in Eq.~(\ref{si-eq:generalcomparison})
uses rate coordinates $(m,d)$.
For the unrestricted response, at least four distinct times make
$x_0,x_1,x_2,x_3$ independent. Put
$b_g=P_{[x_0,x_1,x_2]}^\perp x_3$.
Direct expansion gives
\begin{align}
 \sigma_{\min}(M)
 &\sim\frac{2\sqrt{2}|AC|}{3\sqrt{A^2+C^2}}\|b_g\|d^2,\nonumber\\
 \sigma_{\min}(\mathcal J)&\sim\frac{\sqrt{2}}{3}\|b_g\|d^3 .
 \label{si-eq:generalcomparison}
\end{align}

For the general two-exponential signal, use orthonormal sum and
difference combinations of its two amplitude columns. They
begin with $\sqrt{2}x_0$ and $\sqrt{2}d x_1$.
Their span tends to $\operatorname{span}(x_0,x_1)$.
Writing $a_g=P_{[x_0,x_1]}^\perp x_2$ and using the linear
rate coordinates $(m-d,m+d)$, the projected rate columns have
expansion
\begin{equation}
 d a_g(-A,C)+\frac{d^2}{3}
     P_{[x_0,x_1]}^\perp x_3(-A,-C)+O(d^3).
\end{equation}
Eliminating the leading rank-one part and converting to $(m,d)$
yields the first line of Eq.~(\ref{si-eq:generalcomparison}).
This conversion multiplies both projected-rate singular values
by $\sqrt{2}$.
In the full joint Jacobian, a unit limiting amplitude-difference
direction is accompanied by $O(d)$ rate changes that cancel its
$x_1$ and $x_2$ components. The remaining coefficient is
$-(\sqrt{2}/3)d^3b_g$. The other singular values have orders
$1,1,d$, giving the second line.

For the constrained response the unit limiting null direction
is the $B$ direction, with $\delta d=-d/B+O(d^3)$.
After eliminating the remaining $x_1,x_2$ components its residual
is $-(2/3)d^3b_3$; the other joint singular values have orders
$1,d$. This explains why dividing by the small nuisance column
changes the power seen by the amplitude-projected rate block.
For independently fitted paired amplitudes, the first amplitude
column is $(\Delta v+O(\Delta^2);0)$, and an $O(\Delta)$
change of its rate cancels its leading part. The surviving
component transverse to the other three columns is
$O(\Delta^2)$ with a nonzero coefficient at two distinct positive
times. The full joint minimum is therefore $\Theta(d^2)$.
With supplied $\kappa$, the joint rank is already full at
coincidence, as proved in Sec.~\ref{sec:paired}.
These statements refer to fixed linear amplitude coordinates
and fixed nonzero true amplitudes.

\section{Spherical Stokes resistance and perturbation proof}
\label{si-app:fluid}
For a poloidal Stokes velocity write
$v_r=f(r)Y$, $v_t=g(r)\nabla_\Omega Y$.
Incompressibility gives
$g=(r f'+2f)/\Lambda$.
Regular interior radial powers are $r^{\ell-1},r^{\ell+1}$;
decaying exterior powers are $r^{-\ell},r^{-\ell-2}$.
Matching $f(a)=\dot q_r$ and $g(a)=\dot q_p$ determines the
two coefficients on each side. Substitution into
$\sigma_{rr}=-p+2\eta f'$ and
$\sigma_{rp}=\eta[g'+(f-g)/r]$ gives the resisting traction
coefficients in Eq.~(\ref{eq:fluidmatrix}).
The tangential work carries
$\int|\nabla_\Omega Y|^2d\Omega=\Lambda$.
The work is the positive Stokes dissipation
$2\int_{\rm in}\eta_{\rm in}|D(v)|^2dV+
 2\int_{\rm out}\eta_{\rm out}|D(v)|^2dV$.
For $\ell\ge2$ it vanishes only for zero boundary velocity.
The toroidal powers $r^\ell,r^{-\ell-1}$ similarly yield the
stated toroidal block.

For the perturbation estimates, put
$B_K=K^{1/2}V^{-1/2}$,
$F=(I+V^{-1/2}R_fV^{-1/2})^{-1}$, and
$D_k=V^{-1/2}(\partial_{\theta_k}V)V^{-1/2}$.
At the parameter point under consideration,
$H_f=B_KFB_K^{\mathsf T}$,
$H_0=B_KB_K^{\mathsf T}$, and
\begin{align}
 \|I-F\|_2&=\delta_f,\qquad \|B_K\|_2^2=b,\nonumber\\
 \partial_{\theta_k}H_f&=-B_KFD_kFB_K^{\mathsf T},\nonumber\\
 \partial_{\theta_k}H_0&=-B_KD_kB_K^{\mathsf T}.
\end{align}
These derivative identities differentiate the inverse resistance
with $K,R_f$ fixed. Thus
$\|H_f-H_0\|_2\le\delta_fb$,
$\|\partial_{\theta_k}H_0\|_2\le bg_k$, and
$\|\partial_{\theta_k}H_f-\partial_{\theta_k}H_0\|_2
\le2\delta_fbg_k$.
Both semigroups have norm at most $e^{-\alpha t}$.
Duhamel's integral for their difference gives
Eq.~(\ref{eq:fluidforward}). Differentiating each semigroup
gives the integral of $-E(t-s)(\partial_\theta H)E(s)$.
Replacing its middle factor contributes
$2\delta_fbg_k t e^{-\alpha t}$; replacing its two outer
factors contributes
$\delta_fb^2g_k t^2e^{-\alpha t}$.
Their sum proves Eq.~(\ref{eq:fluidderivative}).

This proof controls additive bulk dissipation with known physical
coefficients in the zero-bending linear model. Changes in stiffness,
including bending elasticity, require their own derivative terms, as do
finite deformation and changes in preparation or observation.
Equation~(\ref{eq:fluidepsilon}) measures bulk dissipation relative
to membrane dissipation on the chosen mode space; small $\epsilon_f$
gives the membrane-dominated limit.

\end{document}